\documentclass[12pt]{iopart}
\usepackage{graphicx} 
\usepackage{bm}
\usepackage{mathrsfs}

\expandafter\let\csname equation*\endcsname\relax
\expandafter\let\csname endequation*\endcsname\relax
\usepackage{amsmath}
\usepackage{lmodern,amssymb,amsthm}
\usepackage{iopams}
\usepackage{xspace}
\usepackage{hyperref}
\usepackage{graphicx}
\usepackage{placeins}
\usepackage{cite}

\usepackage{calc}
\usepackage[utf8]{inputenc}
\usepackage[T1]{fontenc}
\usepackage{verbatim} 

\newcommand{\T}[2]{^{#1}_{\phantom{#1}#2}}
\newcommand{\gG}[5]{\tilde\gamma^{#1#2}\tilde\Gamma^{#3}_{#4#5}}
\newcommand{\scri}{\ensuremath{\mathscr{I}^+}\xspace}

\newcommand{\rh}{r_{\rm h}}
\newcommand{\sh}{\sigma_{\rm h}}
\newcommand{\beq}{\begin{equation}}
\newcommand{\eeq}{\end{equation}}
\newcommand{\bea}{\begin{eqnarray}}
\newcommand{\eea}{\end{eqnarray}}
\renewcommand{\d}[1]{\,\textnormal{d}#1}

\newcommand{\TPI}{\address{Theoretisch-Physikalisches Institut,
    Friedrich-Schiller-Universität Jena,\\ Max-Wien-Platz 1,
          D-07743 Jena, Germany}}

\begin{document}
\title{Initial data for perturbed Kerr black holes on hyperboloidal slices}
\author{David Schinkel, Rodrigo Panosso Macedo, Marcus Ansorg} 
\TPI
\date{\today}

\begin{abstract}
We construct initial data corresponding to a single perturbed Kerr black hole in vacuum. These data are defined on specific hyperboloidal ({\em ``ACMC-''}) slices on which the mean extrinsic curvature $K$ asymptotically approaches a constant at future null infinity \scri. More precisely, we require that $K$ obeys the Taylor expansion $K=K_0 + {\cal O}(\sigma^4)$ where $K_0$ is a constant and $\sigma$  describes a compactified spatial coordinate such that $\scri$ is represented by $\sigma=0$. We excise the singular interior of the black hole and assume a marginally outer trapped surface as inner boundary of the computational domain. The momentum and Hamiltonian constraints are solved by means of pseudo-spectral methods and we find exponential rates of convergence of our numerical solutions. Some physical properties of the initial data are studied with the calculation of the Bondi Mass, together with a multipole decomposition of the horizon. We probe the standard picture of gravitational collapse by assessing 
a family of Penrose-like inequalities and discuss in particular their rigidity aspects. Dynamical evolutions are planned in a future project.
\end{abstract}

\maketitle

\section{Introduction}

One particular objective of numerical relativity is the study of gravitational waves emitted by the interaction of astrophysical objects. The corresponding wave forms permit the construction of templates for gravitational wave detections. Most of these calculations have been carried out on spatially truncated numerical domains. However, as the outgoing gravitational radiation is well-defined only at future null infinity $\scri$, it would be desirable to dispense with finite outer boundaries and corresponding wave extractions obtained through extrapolation techniques.

An elegant alternative approach is the inclusion of $\scri$ as outer boundary of the computational domain, i.e.,~the study of the spacetime on compactified hyperboloidal slices (see \cite{Frauendiener:ConformalInfinity} for a review). This description permits the detailed analysis of the outgoing radiation through the Bondi news function~\cite{Bondi1962,Stewart1989}.

Note that, in this approach, the {\em physical} metric becomes singular at \scri, which gives rise to a decomposition into a {\em conformal} factor, absorbing the singular behavior, and a regular {\em conformal} metric. Friedrich \cite{Friedrich:1983} has formulated a hyperbolic system for the conformal metric on hyperbolidal slices, which is equivalent to the Einstein equations. Unlike the Einstein equations, this system is manifestly regular at $\scri$. The first numerical results of this system (in the context of weak relativistic fields) were obtained by Hübner \cite{Hubner:2000pb}.

Moncrief and Rinne \cite{Moncrief2009} showed that the apparently singular boundary terms appearing in the hyperboloidal concept can be explicitly evaluated at \scri in terms of conformally regular geometric data. As a result of this analysis, they obtained a rather rigidly constrained and gauge fixed formulation of the field equations. The first stable dynamical numerical evolution based on this formulation was published by Rinne \cite{Rinne:2009qx}. The corresponding code provides long-term stable and convergent evolutions of axisymmetrically perturbed Schwarzschild spacetimes.

The construction of initial data as a starting point for a subsequent dynamical evolution of the space time geometry requires the solution of coupled elliptic constraint equations, specifically the Hamiltonian and momentum constraints. In this paper we consider data which are defined on spatially compactified hyperboloidal slices extending up to $\scri$. Previous calculations of such data have been carried out in the context of constant mean curvature ({\em ``CMC-condition''}: mean extrinsic curvature $K={\rm constant}$ throughout the slice). Important contributions were made by Frauendiener \cite{Frauendiener:1998ud} and Buchman et al.~\cite{Buchman:2009ew}. While Frauendiener looked at a particular case of CMC-slices with isotropic extrinsic curvature (that is, the extrinsic curvature is assumed to be proportional to the conformal 3-metric in the slice, which implies constancy of the mean curvature), Buchman et al.~computed data in the context of conformal flatness, taking a generalization of the Bowen-York curvature \cite{Bowen:1980yu} as the solution to the corresponding momentum constraint. 
In these cases, only the Hamiltonian constraint needs to be solved, and it is degenerate at $\scri$ by virtue of the singular physical metric. For isotropic extrinsic curvature, the Hamiltonian constraint reduces to a form which is sometimes called the {\em Yamabe equation}. This equation has been studied extensively and solved numerically with pseudo-spectral methods by Frauendiener \cite{Frauendiener:1998ud}. Andersson et al.~proved conditions under which the Yamabe equation has solutions which extend smoothly up to \scri \cite{Andersson:1992yk}. These criteria were generalized by Andersson and Chru\'sciel \cite{Andersson:1993we,Andersson1994} to the situation in which arbitrary slices obeying the CMC-condition are considered. As an example, these mathematical results apply to the slices used by Buchman et al.~\cite{Buchman:2009ew}.

In this paper, we construct axially symmetric, conformally non-flat initial data corresponding to single perturbed Kerr black holes in vacuum on hyperboloidal slices. We wish, in particular, the data to be smoothly extendible up to \scri. As discussed above, known mathematical criteria regarding this requirement are restricted to CMC-data. It is, however, desirable to dispense with the CMC-condition, as explicitly known black hole CMC-slices are available only for static solutions. The data constructed by Buchman et al.~\cite{Buchman:2009ew} use the Schwarzschild CMC-slices as reference and, as a consequence, possess limited spin parameters of the corresponding black hole. In order to have highly spinning black holes near equilibrium, it is necessary to take the Kerr solution as a starting point. Unfortunately, CMC-slices  of the Kerr solution are known only numerically~\cite{Schinkel:2013tka}, and it is not clear, whether these results can be extended to the situation of binary black hole initial data. Here, we propose a much weaker, mere asymptotic condition ({\em "ACMC"}-condition, see below), whose explicit realization is quite simple and should be straightforward for binary data\footnote{A possible  construction would go about a superposition of two Kerr-Schild-metrics that yields a fictive space-time in which compactified hyperboloidal slices need to be identified. One needs to ensure the freedom of at least three parameters so as to realize the ACMC-condition.}. Most importantly, our numerical results provide a strong indication that the criteria for smooth extensibility of CMC-data up to \scri apply to the ACMC-slices as well. 

In order to approach the constraint equations to be satisfied by the initial data, we consider the conformal transverse traceless decomposition \cite{York1979,York1982}. In this formulation, the momentum constraint contains a term behaving like $\sigma^{-3}\nabla^i K$, where $\sigma$  describes a compactified spatial coordinate such that $\scri $ is represented by $\sigma=0$. Now we can  {\em regularize} the momentum constraint by requiring that the mean curvature $K$ obeys the Taylor expansion $K=K_0 + {\cal O}(\sigma^4)$ where $K_0$ is a constant. Henceforth, we call such slices {\em``ACMC''-slices}, as they fulfill the condition of {\em asymptotically constant mean curvature}. It turns out that the degenerate behavior of the corresponding Hamiltonian constraint is asymptotically the same as that of the Yamabe equation. For this reason, we call this behavior in the following {\em``Yamabe-like.''}

As a starting point for our computations we take the Kerr solution in {\em Kerr coordinates} which are horizon-penetrating. Using the {\em ``height-function''-technique} from Zenginoğlu \cite{Zenginoglu:2007jw}, we apply a particular coordinate transformation in order to obtain compactified hyperboloidal ACMC-slices. This is to say, the coordinate transformation is chosen such that, in these compactified slices, the extrinsic mean curvature $K$ is asymptotically constant, $K=K_0 + {\cal O}(\sigma^4)$. Now we consider the momentum constraint as well as the Yamabe-like Hamiltonian constraint that arises in the conformal transverse traceless approach. The corresponding computational domain is bounded by $\scri$ (outer boundary) and by a marginally outer-trapped surface which serves as inner boundary. In other words, we excise the black hole interior and require that at the excision surface, the expansion of outgoing null rays vanishes. Of course, for the Kerr solution we started with, the constraints are identically satisfied. It is, however, easy to modify the set-up slightly and to compute initial data which correspond to {\em perturbed} Kerr black holes. To be more specific, in the conformal transverse traceless approach, the constraint equations are written in terms of freely specifiable data. We keep these data, as well as the boundary values of the solution to the momentum constraint as they are given by the Kerr solution in our ACMC-slices, but modify the {\em coordinate location} of the marginally outer-trapped surface. We then solve the constraints in the new computational domain given by the modified inner boundary and obtain perturbed Kerr black hole data. 

These initial data have been computed with a single domain pseudo-spectral method, which provides us with a rapid exponential convergence rate of the numerical solutions. This exponential fall-off is a strong indication that for the ACMC-slices the same criteria, regarding the smooth extensibility of the data up to \scri, hold as for strict CMC-slices (as mentioned above). Unfortunately, rigorous mathematical results on the smoothness of non-CMC-data are rare. Isenberg and Park~\cite{Isenberg97} discussed the existence of so-called {\em asymptotically hyperbolic initial data}. Existence and boundedness of solutions belonging to another class of non-CMC-data, for which a certain {\em limit equation} admits no non-trivial solutions, are investigated in~\cite{Gicquaud12}. Interesting results on the existence of non-CMC-data in the context of apparent horizon boundary conditions were presented in~\cite{Dilts:2013uea,Holst:2014kva}. However, the several types of data studied in those articles should be distinguished from the solutions treated in this paper, as the  ACMC-condition, $K=K_0 + {\cal O}(\sigma^4)$, is quite different from the assumptions made therein. In fact, in view of our numerical results it would be interesting to study the applicability of the mathematical methods presented in \cite{Andersson:1993we,Andersson1994} to the ACMC-condition \cite{Andersson2013}.

In a post-processing analysis, we study some physical properties of our initial data. We first use an apparent horizon finder to check whether the numerical inner boundary corresponds, in fact, to the apparent horizon, i.e., whether our numerical boundary is the {\em outermost} marginally outer-trapped surface. As expected, this is the case for small perturbations around the original Kerr solution. However, stronger perturbations show a richer structure, with the formation of an apparent horizon that encloses the numerical inner boundary. We determine the deviation from the (unperturbed) Kerr solution by the study of mass and angular momentum multipoles that characterize the marginally outer-trapped surface. We find distinct differences with respect to the corresponding Kerr expressions, which confirms that our data describe perturbed rotating black holes (and not other different slices of the Kerr space-time). Finally, we compute the Bondi Mass of our slice and, together with the black hole angular momentum, we look at some aspects in the standard picture of gravitational collapse (see \cite{Jaramillo:2011} for an overview). Specifically we probe some black hole inequalities that were conjectured with the assumption of the Weak Cosmic Censorship. In particular, Dain et al.~generalized in \cite{Dain2002} the Penrose inequality~\cite{Penrose73} in a way to include the black hole angular momentum. We extend their work and demonstrate that, for our results, a stronger inequality for black hole space-times in terms of the Bondi Mass is valid.

The paper is organized as follows. In section \ref{sec:ACMC} we construct hyperboloidal ACMC-slices of the Kerr solution. Next, in section \ref{sec:constraint_equatins_compactified_hyperboloidal_slices} we discuss the constraint equations within the conformal transverse traceless decomposition and elaborate on regularity criteria valid at \scri. The inner boundary is the subject of section \ref{sec:MOTS}, while we present the conditioning as well as the description of the numerical treatment in section \ref{sec:Numeric}. In section \ref{sec:Results}, we provide exemplary results corresponding to strongly perturbed Kerr data as well as to weakly perturbed data with large rotation parameter. The physical properties of our initial data are presented in section \ref{sec:Physical}. We conclude in section \ref{sec:Discussion} with a discussion and a perspective of future work. The appendix contains the explicit form of the auxiliary function $A$ used in the coordinate transformation to obtain ACMC-slices (see \ref{sec:Appendix_A}), and a comparison between CMC-~and ACMC-slices in the Schwarzschild case (see \ref{sec:ACMC_CMC_Comparison}).

Note that we use the following conventions. Greek indices run from 0 to 3 where 0 denotes the time component. Small Latin indices are from 1 to 3, covering spatial indices. Conformal (unphysical) objects will be decorated with a tilde. We use units in which the speed of light as well as Newton's constant of gravitation are unity.

\section{The Kerr solution in ACMC-slices}\label{sec:ACMC}
As mentioned above, the construction of initial data pursued in this paper is based on the conformal transverse traceless decomposition of the constraints on compactified hyperboloidal slices. In order to regularize the momentum constraint in this formulation, we consider ACMC-slices on which the mean curvature $K$ obeys the Taylor expansion $K=K_0+{\cal O}(\sigma^ 4)$. In the following, we identify such slices within the Kerr solution, which then will serve as a starting point for the generation of perturbed Kerr data.

We begin with the horizon-penetrating line element in Kerr coordinates $(V,r,\theta,\varphi)$,
\begin{multline}
 ds^2=-\left(1-\frac{2M r}{\rho ^2}\right)dV^2 + 2dV dr -\frac{4M r a}{\rho ^2}\sin^2\theta\, dV d\varphi \\ + \rho ^2d\theta^2 - 2a \sin^2\theta\,drd\varphi + \frac{1}{\rho ^2}\left[\left(r^2+a^2\right)^2 - \Delta  a^2\sin^2\theta\right]\sin^2\theta\,d\varphi^2.
 \label{eqn:KerrCoordinates}
\end{multline}
Here we have
\beq
 \rho =\sqrt{r^2+a^2\cos^2\theta}, \qquad \Delta =r^2-2M r+a^2,
\eeq
where $M$ and $J=aM$ are,  respectively, the mass and angular momentum of the black hole. Note that in the following we use the dimensionless angular momentum parameter \[j=\dfrac{a}{M}=\dfrac{J}{M^2}.\]
 
The event horizon of the black hole is located at $r=\rh=M(1+\sqrt{1-j^2})$ which runs from $\rh=2M$ in the Schwarzschild case $(j=0)$ to $\rh=M$ for the extreme Kerr black hole $(|j|=1)$.

We introduce hyperboloidal slices through a coordinate transformation that is motivated by an asymptotic integration of outgoing radial null rays, defined by
\[d s^2=0,\quad d\theta=0,\quad d\varphi=0,\quad dVdr>0.\]
The corresponding equation
\beq
dV=\frac{2 dr}{1-\dfrac{2Mr}{\rho^2}}=2\left[1+\frac{2M}{r}+{\cal O}(r^{-2})\right]dr \label{eq:radial_null_rays}
\eeq
is solved asymptotically by
\[V=4M\left[{\rm constant} + \left(\frac{r}{2M} + \log\frac{r}{2M}\right)\right].\]
From this form, we obtain a coordinate transformation which leads us from Kerr coordinates to general axially symmetric, compactified hyperboloidal slices by the {\em ansatz}\footnote{Note that through this ansatz we retain the horizon penetrating nature of the coordinates.}
\begin{eqnarray}
r &=& \frac{2M}{\sigma}\label{eq:CoordTrafo_r}\\
V &=& 4M\Bigg[\tau + \underbrace{\left(\frac{1}{\sigma} - \log(\sigma)  + A(\sigma,\cos\theta) \right)}_{\displaystyle h}\Bigg]\label{eq:CoordTrafo_V}.
\end{eqnarray}
In the new coordinates $(\tau,\sigma,\theta,\varphi)$, the compactified hyperboloidal slices are described by $\tau={\rm constant}$. The coordinate location of future null infinity $\scri$ is given by $\sigma=0$ while the  event horizon is placed at \begin{equation}
\label{eq:sh}
\sigma=\sh=\frac{2}{1+\sqrt{1-j^2}}.
\end{equation}
The coefficients $g_{\mu\nu}$ of the physical metric can be written in the form 
\beq
g_{\mu\nu}=\Omega^{-2}\tilde g_{\mu\nu}.\label{eq:conformal_decomposition}
\eeq
Here, $\tilde g_{\mu\nu}$ denotes a conformal metric, and $\Omega$ is a {\em conformal factor} that vanishes linearly at $\scri $, i.e.~$\Omega|_{\scri}=0,\,\,  \partial_\sigma\Omega|_{\scri}\ne 0$. For the function $A$ appearing in (\ref{eq:CoordTrafo_V}), an arbitrary smooth expression in terms 
of $\sigma\in[0,\sigma_{\rm h}]$ and \[\mu:=\cos\theta\in[-1,1]\] can be chosen\footnote{The requirement of axisymmetry implies that $A$ has to depend on $\cos\theta$, which we denote henceforth by $\mu$.}, as long as one ensures that the resulting hyperboloidal slices $\tau={\rm constant}$ are spacelike everywhere away from $\scri$. This construction of hyperboloidal slices is equivalent to the {\em height function technique} \cite{Zenginoglu:2007jw}, and the corresponding height function $h$ is indicated in (\ref{eq:CoordTrafo_V}). Along these lines, several formulations have been identified that provide hyperboloidal slices of the Kerr solution, see~\cite{Racz:2011qu,Jasiulek:2011ce}\footnote{Note that for the hyperboloidal slices used in \cite{Racz:2011qu,Jasiulek:2011ce}, no specific requirements in regard of the asymptotics of the mean extrinsic curvature $K$ were specified.}. In the following we make use of the freedom to choose the smooth function $A$ in order to generate slices that satisfy the ACMC-condition 
\beq
K=K_0+{\cal O}(\sigma^4),\qquad K_0={\rm constant.}\label{eq:ACMC_Kerr}
\eeq
Here $K$ is the {\em mean curvature}
\beq
K=K_{\mu\nu}g^{\mu\nu},
\eeq
i.e.,~the trace of the {\em extrinsic curvature} $K_{\mu\nu}$,
\beq
K_{\mu\nu}=\frac{1}{2}\mathcal{L}_ng_{\mu\nu}.
\eeq
In this definition, $\mathcal{L}_n$ is the Lie derivative along the future pointing unit vector $n^\mu$ normal to the slices $\tau={\rm constant.}$ We follow the sign convention employed by Wald \cite{Wald:1984}, which was also used in \cite{Buchman:2009ew}. In this convention (opposite to the widely adopted one by Misner, Thorne \& Wheeler \cite{MTW:1973}), the mean curvature $K$ assumes {\em positive} values at \scri. \\
Now, for a given function $A(\sigma,\mu)$, the Taylor expansion of $K$ with respect to $\sigma$ at \scri ($\sigma=0$), considered in a slice $\tau={\rm constant}$, has the following structure:
\beq
K=K_0 + K_1\sigma^1+K_2\sigma^2+K_3\sigma^3 +O(\sigma^4)
\eeq 
with rather longish expressions for the coefficients $K_j=K_j(\mu)$. 
Putting 
\[
		K_0=\text{constant},\qquad K_1=K_2=K_3\equiv0,
\]
 we obtain asymptotic conditions for $A(\sigma,\mu)$,
\begin{eqnarray}
\label{eq:A0}  A\Big|_{\sigma=0}&=&a_0 \label{eq:A0} \\
\label{eq:A1}  \partial_\sigma A\Big|_{\sigma=0}&=&a_1 - \frac{j^2\mu^2}{16}\\
\label{eq:A2}  \partial^2_\sigma A\Big|_{\sigma=0}&=&-1 + \frac{j^2}{4}\\
\label{eq:A3}  \partial^3_\sigma A\Big|_{\sigma=0}&=&-2(4 +6 a_1 +3a_1^2) \\&&+ j^2\left(1 +\frac{3}{4}\mu^2(1+a_1) - \frac{3}{128}j^2\mu^4 \right),\nonumber
\end{eqnarray}
with $a_0$ being a linear function of $\mu$. As such an angular dependence would introduce a coordinate asymmetry regarding the equatorial plane, we choose $a_0$ to be a constant, which, in turn, implies that $a_1$ also becomes constant. A detailed derivation of eqs.~(\ref{eq:A0}-\ref{eq:A3}) can be found in~\cite{Schinkel:2013tka}.

As mentioned above, we need to ensure that the corresponding slices $\tau={\rm constant}$ are spacelike, which means that the conformal lapse \[\tilde\alpha=\Omega\alpha=\frac{1}{\sqrt{-\tilde{g}^{00}}}\] is well defined everywhere. Now, for numerical reasons, it would be desirable to have a ``very'' smooth function $\tilde\alpha$, meaning that a spectral expansion with respect to Chebyshev polynomials shows a rapid exponential fall-off of the corresponding coefficients $|c^{(\tilde\alpha)}_k|$\,\footnote{\label{fn:Cheb_alpha} As a representative, we consider $\tilde\alpha(\sigma,\mu=0)$, i.e.~the conformal lapse as a function of $\sigma\in[0,\sigma_{\rm h}]$ along the equatorial plane $\theta=\pi/2$.}.  Indeed we find that a rapid convergence rate of these coefficients is a good indication of high numerical accuracy of the results obtained from the subsequent pseudo-spectral solution of the constraints. We thus aim at a choice for the function $A$ such that on the one hand the conditions (\ref{eq:A0})-(\ref{eq:A3}) are satisfied and on the other hand the $|c^{(\tilde\alpha)}_k|$ fall off rapidly.

We have found that these requirements are met by an appropriate diagonal Pad\'e expansion. In this approach, the coefficients $b_k=b_k(\mu)$ and $c_k=c_k(\mu)$ of the rational function
\beq
\mathfrak{P}_n(\sigma,\mu)=\frac{\sum\limits_{k=0}^nb_k\sigma^k}{1+\sum\limits_{k=1}^nc_k\sigma^k}
\eeq
are determined by the first $2n$ derivatives of $A$ at $\sigma=0$, i.e.~through the conditions given in (\ref{eq:A0})-(\ref{eq:A3}). Since $a_0$ represents an irrelevant time shift, we can set it to zero, which implies $b_0=0$. In order to satisfy the remaining three conditions (\ref{eq:A1})-(\ref{eq:A3}), we need at least a second order Pad\'e expansion, $n=2$, which provides us with the freedom to choose the fourth derivative 
\[a_4=\partial^4_\sigma A\Big|_{\sigma=0}.\] 
For simplicity, we set $a_4$ to be a constant. In this way, our Pad\'e approximation contains two undetermined parameters, $a_1$ and $a_4$, and we can use this freedom to obtain rapid fall-off rates of the corresponding spectral coefficients $c^{(\tilde\alpha)}_k$. A satisfactory choice is given by 
\begin{eqnarray}
a_1&=&\frac{113}{100} + \frac{9}{500} |j| + \frac{3}{40} j^2 \label{eq:a1_j}\\[3mm]
a_4&=&4! \,\,\left(\frac{161}{10} + \frac{3}{4} |j| - \frac{5}{2} j^2\right),\label{eq:a4_j}
\end{eqnarray}
which describes the two constants parametrically in terms of the dimensionless angular momentum $j$. The explicit form of the resulting function \beq A(\sigma,\mu)=\mathfrak{P}_2(\sigma,\mu)\eeq
as well as a discussion regarding the derivation of the choice (\ref{eq:a1_j},\ref{eq:a4_j}) can be found in \ref{sec:Appendix_A}. Note that the corresponding mean curvature $K$ assumes the required ACMC-structure,
\beq
K(\sigma,\mu)=\frac{3}{M\sqrt{16 (1+a_1)-j^2}} + O(\sigma^4). \label{eq:aCMC_K_Taylor}
\eeq

\section{The constraints in the conformal transverse traceless decomposition on compactified hyperboloidal ACMC-slices}\label{sec:constraint_equatins_compactified_hyperboloidal_slices}

\subsection{Constraint equations}
In the standard 3+1 decomposition of spacetime, the line element is written as
\beq
ds^2=-\alpha^2 dt^2 + \gamma_{ij}\left(dx^i + \beta^i dt\right)\left(dx^j + \beta^j dt\right).
\eeq
with $\alpha$ being the {\em lapse} function, $\beta^i$ the {\em shift vector}, and $\gamma_{ij}$ the induced three-metric of the spatial slice.
The vacuum constraint equations valid in that slice are the momentum constraints
\beq
 \nabla_{i}(K^{ij} - \gamma^{ij}K)=0
\eeq
and the Hamiltonian constraint
\beq
R + K^2 - K_{ij}K^{ij}=0.
\eeq
Here, $\nabla_i$ is the covariant derivative associated with the induced metric $\gamma_{ij}$, and $R$ is the associated curvature scalar, while the $K_{ij}$ denote the spatial components of the extrinsic curvature. 

There are several formulations for decomposing the constraint equations. Here we use the {\em conformal transverse traceless decomposition} (CTT) \cite{York1979, York1982}, in which we write
\beq
\begin{array}{ccc}
 K_{ij}&=&\Omega\tilde A_{ij} + \dfrac{1}{3\Omega^2}\tilde\gamma_{ij}K,\\[5mm]
 K^{ij}&=&\Omega^5\tilde A^{ij} + \dfrac{\Omega^2}{3}\tilde\gamma^{ij}K 
\end{array}
 \label{eq:decomposition_extrinsic_curvature}
\eeq
with the conformal three-metric defined as
\[\tilde\gamma_{ij}=\Omega^2\gamma_{ij},\qquad \tilde\gamma^{ij}=\Omega^{-2}\gamma^{ij},\]
which is the metric to raise and lower indices of conformal quantities, and the traceless part $\tilde A^{ij}$,
\[\tilde\gamma_{ij}\tilde A^{ij}=0=\tilde\gamma^{ij}\tilde A_{ij}. \]
We further decompose $\tilde A^{ij}$ as
\beq
\tilde{A}^{ij}=(\tilde{\mathcal{L}}V)^{ij} + M^{ij},\label{eq:decomposition_A}
\eeq
where 
\[(\tilde{\mathcal{L}}V)^{ij}=\tilde\nabla^iV^j+\tilde\nabla^jV^i-\frac{2}{3}\tilde\gamma^{ij}\tilde\nabla_kV^k\]
and $\tilde{\nabla}_i$ denotes the covariant derivative with respect to $\tilde{\gamma}_{ij}$.


In the CTT approach, the symmetric tracefree tensor $M^{ij}$, the mean curvature $K$, and the conformal metric $\tilde{\gamma}_{ij}$ are considered as freely specifiable data. Given these, the momentum constraint becomes
\beq
\tilde{\Delta}_\mathcal{L}V^{i}=\frac{2}{3\Omega^3}\tilde{\nabla}^i K - \tilde{\nabla}_jM^{ij} \label{eq:momentum_constraint_conformal_prestep}
\eeq
with
\[\tilde{\Delta}_\mathcal{L}V^{i}=\tilde{\nabla}_j(\tilde{\mathcal{L}}V)^{ij}.\]
Here we have followed the reasoning of the CTT-formulation presented in~\cite{Cook:2000lr} which, by virtue of the decomposition (\ref{eq:decomposition_A}), eventually arrives at the constraint equation (\ref{eq:momentum_constraint_conformal_prestep}) where $M^{ij}$ is  symmetric tracefree but not necessarily transverse.
Equation (\ref{eq:momentum_constraint_conformal_prestep}) is to be considered subject to boundary conditions, imposed here at \scri and at an inner marginally outer-trapped surface boundary (see below). Note that two different sets of curvature quantities, $\{\stackrel{1}{M}\!{}^{ij}, \stackrel{1}{V}\!{}^{i}\}$ and $\{\stackrel{2}{M}\!{}^{ij}, \stackrel{1}{V}\!{}^{i}\}$, which are related through
\[
\stackrel{1}{M}\!{}^{ij}-\stackrel{2}{M}\!{}^{ij}=(\tilde{\mathcal{L}}W)^{ij},\qquad \stackrel{1}{V}\!{}^{i}-\stackrel{2}{V}\!{}^{i}=-W^i
\]
for some vector field $W^i$, will lead to the same extrinsic curvature $K_{ij}$.

We now turn our attention to the Hamiltonian constraint, which we rewrite by utilizing the decomposition (see, e.g., \cite{Frauendiener:ConformalInfinity})
\beq
\Omega=\omega\phi^{-2}, \label{eq:decomposition_Omega}
\eeq
where $\omega$ is a prescribed spatial function\footnote{In this paper, we choose $\omega$ to always
agree with the compactified radial coordinate $\sigma$.} which describes \scri by $\omega|_{\scri}=0$. Note that $\omega$ has similar properties to $\Omega$, i.e.,~it is positive away from $\scri$ and possesses non-vanishing gradient at \scri. The auxiliary potential $\phi$ is chosen to be positive everywhere. 

The Hamiltonian constraint then turns into
\begin{multline}
 - \left(\omega^2 \tilde{R} + 4\omega\tilde{\nabla}^2\omega - 6\tilde{\nabla}^a\omega\tilde{\nabla}_a\omega \right)\phi - 8\omega\tilde{\nabla}^a\phi\tilde{\nabla}_a\omega \\ +  8\omega^2\tilde{\nabla}^2\phi = \frac{2}{3}K^2\phi^5 - \omega^6 \tilde{A}_{ij}\tilde{A}^{ij}\phi^{-7}.\label{eq:Hamilton_constraint_conformal}
\end{multline}
Here $\tilde{R}$ denotes the curvature scalar associated with $\tilde{\gamma}_{ij}$.  Note that for isotropic extrinsic curvature, $\tilde{A}_{ij}\equiv 0$, the last term vanishes, and (\ref{eq:Hamilton_constraint_conformal}) turns into the   {\em Yamabe equation}\footnote{Isotropic extrinsic curvature implies $K={\rm constant}$, i.e.,~the corresponding slices satisfy the CMC-condition. Note that there are typos in equation (34) of \cite{Frauendiener:ConformalInfinity} and equation (4) of \cite{Frauendiener:1998ud}.}.

Finally, with the decomposition of the conformal factor (\ref{eq:decomposition_Omega}), the momentum constraints read as
\beq
\tilde{\Delta}_\mathcal{L}V^{i}=\frac{2}{3}\frac{\phi^6}{\omega^3}\tilde{\nabla}^i K - \tilde{\nabla}_jM^{ij}.\label{eq:momentum_constraint_conformal}
\eeq
In sum, the determination of valid initial data on compactified hyperboloidal slices in the CTT approach consists of the solution of the  conformal constraint system (\ref{eq:Hamilton_constraint_conformal}, \ref{eq:momentum_constraint_conformal}) for the potentials $\phi$ and $V^i$. As mentioned above, $M^{ij}$, $K$, $\tilde{\gamma}_{ij}$ as well as $\omega$ are prescribed quantities, and $\tilde{A}^{ij}$ is given in terms of $V^i$ and $M^{ij}$ by equation (\ref{eq:decomposition_A}).

\subsection{Regularity of the constraint equations}{\label{subsec:RegCond}}
The momentum constraint (\ref{eq:momentum_constraint_conformal}) contains a term \mbox{$\sim\Omega^{-3}\tilde{\nabla}^iK$} which is bounded up to \scri only when the mean curvature satisfies the ACMC-condition
\beq
 K=K_0 + K_4\omega^4,\qquad K_0={\rm constant}.
\eeq
If the global CMC-condition is imposed, that is $K_4\equiv 0$, then the constraints decouple, i.e.,~at first, the momentum constraints can be solved for $V^i$ and then the Hamiltonian constraint for $\phi$. Although we lose this separability for $K_4\not\equiv 0$, we retain the structure of the constraint equations at \scri. To be more specific, the momentum constraint is a non-singular elliptic system of differential equations, requiring the prescription of inner and outer boundary conditions for a unique solution. At the same time, the Hamiltonian constraint assumes a {\em Yamabe-like} structure, i.e.~it looks asymptotically up to order $\omega^4$ like the {\em Yamabe equation}. If we make the {\em ansatz}\footnote{The so-called {\em little-o} notation, $f(\delta) = o(g(\delta))$ as $\delta\to 0$, means that for every positive constant $\epsilon$ 
there exists a constant $\delta_\epsilon$ such that $|f(\delta)|\leq\epsilon|g(\delta)|$ for all $\delta<\delta_\epsilon$.}
\beq\label{eq:Taylor_phi}
\phi = \phi_0 + \phi_1\omega + \phi_2\omega^2 + o(\omega^2),\quad
\partial_\omega\phi = \phi_1 + 2\phi_2\omega + o(\omega),\quad
\partial^2_\omega\phi = 2\phi_2 + o(1),
\eeq
where we assume the quantities $\{\phi_0, \phi_1, \phi_2\}$ to be bounded, then we find by virtue of the Yamabe-like structure, that these three terms can be constructed explicitly, e.g.,~for $\phi_0$ we obtain:
\beq
\phi_0=\sqrt[4]{9K_0^{-2}(\tilde\nabla^a\omega\tilde\nabla_a\omega)|_{\mathscr{I}^+}}.\label{eq:taylor_hamilton_constraint}
\eeq
However, in a corresponding ansatz with terms higher than quadratic order, these higher order terms would not be determinable {\em a priori}. In our numerical treatment we impose (\ref{eq:taylor_hamilton_constraint}) as a Dirichlet boundary condition, see discussion in Section \ref{sec:Numeric}. Note that, on the other hand, the quantities $\phi_1$ and $\phi_2$ need not be prescribed inside the numerical scheme.

Andersson and Chru\'sciel investigated criteria for smooth solutions to the Hamiltonian constraint on hyperboloidal slices satisfying the CMC-condition \cite{Andersson:1993we,Andersson1994}. They found that $\phi$ (and hence $\Omega$) is smooth in a vicinity of \scri provided that at \scri, the conformal extrinsic curvature 
\beq
\tilde\kappa_{ij}=\frac{1}{2}{\cal L}_{\tilde s}\tilde q_{ij}
\label{eq:kappa}
\eeq
is proportional to the induced conformal metric on \scri
\beq
\tilde q_{ij}=\tilde\gamma_{ij}-\tilde s_i\tilde s_j,\label{eq:q_ij}
\eeq
that is, that there is a scalar $\lambda$ defined on \scri such that
\beq
\left(\tilde \kappa_{ij}-\lambda \tilde q_{ij}\right)|_{\scri}=0.\label{eq:extrinsic_curvature_scri}
\eeq
Here, $\tilde{s}^i$ is the conformal unit vector normal to \scri in the hyperboloidal slice. The condition (\ref{eq:extrinsic_curvature_scri}) is equivalent to requiring that \scri be {\em shear-free}. 

In this work we assume that the shear-freeness of \scri (\ref{eq:extrinsic_curvature_scri}) ensures likewise the smoothness of solutions to the constraint system (\ref{eq:Hamilton_constraint_conformal}, \ref{eq:momentum_constraint_conformal}) which arises in the context of the more general ACMC-condition. As we use for the conformal metric $\tilde\gamma_{ij}$ the Kerr expression in ACMC-slices (constructed in section \ref{sec:ACMC}), we can explicitly check the validity of  (\ref{eq:extrinsic_curvature_scri}). We find $\lambda\equiv 0$, i.e.,~the condition is fulfilled in a trivial manner.

Note that for most general CMC-slices a Taylor expansion of the auxiliary potential $\phi$ with respect to $\omega$ (i.e.~taken in the normal direction with respect to \scri) cannot be carried out but breaks down at 3rd order in $\omega$ (cf.~(\ref{eq:Taylor_phi})). To be more specific, the analysis reveals terms $\propto\omega^3\log(\omega)$. These logarithmic terms vanish if the shear-freeness of \scri [i.e.,~Eq.~(\ref{eq:extrinsic_curvature_scri})]
is imposed. Bardeen et al.~\cite{Bardeen:2011ip} discuss the fact that this condition is equivalent to the absence of outgoing radiation at \scri. Here we are interested in smooth initial data which do not possess these logarithmic terms. Following the arguments presented in \cite{Bardeen:2011ip} and assuming their applicability to our ACMC-slices, we expect that the data constructed below do not carry outgoing radiation at $\scri$.

\section{Marginally outer-trapped surfaces as inner boundaries}
\label{sec:MOTS}
In this paper, we compute initial data corresponding to a perturbed Kerr black hole in vacuum. In order to describe the black hole, we excise its singular interior and require the excision boundary ${\cal H}$ to be a marginally outer-trapped surface \cite{Jaramillo:2011,Penrose65}. We thus obtain an inner boundary of our computational domain.

In order to obtain the boundary conditions on the field $\phi$, let us first consider a closed 2-surface ${\cal S}$ embedded into the 3-slices. One can construct the null vectors
\[ 
\ell^\pm_\mu = \frac{1}{\sqrt{2}}(n_\mu \pm s_\mu),\] with $n^\mu$  the future pointing unit normal to the 3-slices $\tau=$ constant and $s^\mu$ the outward pointing unit vector normal to the surface ${\cal S}$. The corresponding expansions scalars $\Theta_\pm = q^{\mu \nu}\nabla_\mu \ell^\pm_\nu$ (with $q^{\mu \nu}$ the induced metric on ${\cal S}$) read 
\bea
\Theta_{\pm} & = & \frac{1}{\sqrt{2}} \left( K - s^a s^b K_{a b} \pm \nabla_a s^a \right) \nonumber \\
& = &  \frac{1}{\sqrt{2}} \left[ \frac{2}{3}K - \Omega^3\tilde{s}^a \tilde{s}^b \tilde{A}_{a b}  \pm  \left( \Omega \tilde\nabla_a \tilde{s}^a -2\tilde{s}^a\partial_a \Omega \right)  \right]. \label{eq:Expansions}
\eea
The second line in (\ref{eq:Expansions}) was written in terms of the conformal quantities and $\tilde{s}^i=\Omega^{-1}s^i$ was introduced. It takes into account the decomposition of the extrinsic curvature given by eq. (\ref{eq:decomposition_extrinsic_curvature}). 

For a marginally outer-trapped surface, the expansion scalar for outgoing null rays $\Theta_+$  vanishes. In ACMC-slices of the Kerr solution (see section \ref{sec:ACMC}), this condition is satisfied at the location of the event horizon $r=\rh$, corresponding to $\sigma=\sigma_{\rm h}=2/(1+\sqrt{1-j^2})$. Furthermore, this surface is the {\em outermost} marginally trapped surface, i.e.,~the {\em apparent horizon} \cite{Huq2002a}. Also notice that at the outer numerical boundary, i.e.~at $\scri$, the condition $\Theta_-=0$ is realized\footnote[1]{To get $\Theta_-=0$ at $\scri$, one computes $\Theta_-$ via (\ref{eq:Expansions}) for 2-surfaces $\sigma=\sigma_0={\rm constant}\ne 0$ and determines the limit as $\sigma_0$ goes to zero.} 
as a direct consequence of the condition (\ref{eq:taylor_hamilton_constraint}).

Writing $\Theta_+ = 0$ in (\ref{eq:Expansions}) with the help of (\ref{eq:decomposition_Omega}), we get an inner boundary condition valid at ${\cal H}$, containing $\phi$ as well as its first derivatives,
\beq
\tilde{s}^{a}\partial_a\phi + \frac{\phi}{4}\left(\tilde{\nabla}_a\tilde{s}^a-\frac{2\tilde{s}^a\partial_a\omega}{\omega}\right) + \frac{K\phi^3}{6\omega} - \frac{\omega^2}{4\phi^3}\tilde{s}^{a}\tilde{s}^{b}\tilde{A}_{ab}=0\label{eq:inner_boundary_conformal}.
\eeq

For the construction of perturbed Kerr data below, we require this condition at a coordinate location $\sigma_{\cal H}$ that differs from the Kerr value $\sigma_{\rm h}$ [see Eq.~(\ref{eq:sh})]. At the same time, we keep all coefficients in the constraint equations as well as the boundary conditions of the vector $V^i$ in the form that results from the Kerr solution for specifically prescribed rotation parameter $j$ in ACMC-slices. 

If the perturbation is chosen to be sufficiently small, we ensure that ${\cal H}$ still describes the outermost marginally outer-trapped surface~\cite{Andersson:2007gy,Andersson:2008up}, i.e.,~the apparent horizon. However, as will be described in section \ref{sec:HorizonFinder}, strong perturbations lead to the formation of new trapped surfaces.

\section{Numerical solution of the constraint equations}\label{sec:Numeric}

In this section we explicitly describe the steps in the set-up of the constraints and their numerical solution. 
Using the coordinates $(x^0,x^1,x^2,x^3)=(\tau,\sigma,\mu,\varphi)$ and putting $\omega=\sigma$, we write the Hamiltonian constraint as
\beq
\label{eq:Num:Hamilt_Constraint}
H_1\phi_{,11} + H_2\phi_{,22} + H_3\phi_{,12} + H_4\phi_{,1} + H_5\phi_{,2} + H_6\phi \\+ H_7\phi^5 + H_8\phi^{-7}=0,
\eeq
where the $H_j$ are defined as:
\begin{eqnarray}
H_1&=&8\sigma^2\tilde\gamma^{11}\nonumber\\
H_2&=&8\sigma^2\tilde\gamma^{22}\nonumber\\
H_3&=&16\sigma^2\tilde\gamma^{12}\nonumber\\
H_4&=&-8\sigma^2\tilde\gamma^{ab}\tilde\Gamma^{1}_{ab}-8\sigma\tilde\gamma^{11}\nonumber\\
H_5&=&-8\sigma^2\tilde\gamma^{ab}\tilde\Gamma^{2}_{ab}-8\sigma\tilde\gamma^{2\,1}\\
H_6&=&-\sigma^2\tilde R + 4\sigma\tilde\gamma^{ab}\tilde\Gamma^{1}_{ab} + 6\tilde\gamma^{11}\nonumber\\
H_7&=&-\frac{2}{3}K^2\nonumber\\
H_8&=&\sigma^6(\mathfrak{k}\T{ij\,b}{a}V\T{a}{,b} + \mathfrak{l}\T{ij}{a}V^a + \mathfrak{m}^{ij})(\mathfrak{k}_{ij\;\;\,a}^{\;\;\;\;b} V\T{a}{,b} + \mathfrak{l}_{ija}V^a + \mathfrak{m}_{ij}).\nonumber
\end{eqnarray}
By virtue of the decomposition (\ref{eq:decomposition_A}) we have written here $\tilde A^{ij}$ and $\tilde A_{ij}$ as
\beq
\begin{array}{ccc}
\eqalign{
\tilde{A}^{ij}=\mathfrak{k}\T{ijb}{a}V\T{a}{,b} + \mathfrak{l}\T{ij}{a}V^a + \mathfrak{m}^{ij}\\
\tilde{A}_{ij}=\mathfrak{k}_{ij\phantom{b}a}^{\phantom{ij}b} V\T{a}{,b} + \mathfrak{l}_{ija}V^a + \mathfrak{m}_{ij},}
\end{array}
 \label{eq:Numerik_Add}
\eeq
where the $\mathfrak{k}$, $\mathfrak{l}$, $\mathfrak{m}$ are given by 
\begin{eqnarray}
\mathfrak{k}\T{ijb}{a}&=&\tilde\gamma^{bi}\delta^j_a + \tilde\gamma^{bj}\delta^i_a -\frac{2}{3}\tilde\gamma^{ij}\delta^b_a\nonumber\\
\mathfrak{l}\T{ij}{a}&=&\left( \tilde\gamma^{ki}\tilde\Gamma^{j}_{ka} + \tilde\gamma^{kj}\tilde\Gamma^{i}_{ka} - \frac{2}{3}\tilde\gamma^{ij}\tilde\Gamma^{k}_{ka} \right)\\
\mathfrak{m}^{ij}&=&M^{ij}\nonumber
\end{eqnarray}
and
\begin{eqnarray}
\mathfrak{k}_{ij\phantom{b}a}^{\phantom{ij}b}&=&\tilde\gamma_{aj}\delta^b_i + \tilde\gamma_{ia}\delta^b_j -\frac{2}{3}\tilde\gamma_{ij}\delta^b_a\nonumber\\
\mathfrak{l}_{ija}&=&\left( \tilde\gamma_{kj}\tilde\Gamma^{k}_{ia} + \tilde\gamma_{ik}\tilde\Gamma^{k}_{ja} - \frac{2}{3}\tilde\gamma_{ij}\tilde\Gamma^{k}_{ka} \right)\\
\mathfrak{m}_{ij}&=&M_{ij}\label{eq:m_of_M}.\nonumber
\end{eqnarray}

The momentum constraints read
\beq\label{eq:Num:Mom_Constraint}
\mathfrak{a}^{irq}_{\phantom{irq}p}V^{p}_{\phantom{p},qr} + \mathfrak{b}^{iq}_{\phantom{iq}p}V^{p}_{\phantom{p},q}+\mathfrak{c}^{i}_{\phantom{i}p}V^{p} + \mathfrak{d}^i + \mathfrak{e}^i\phi^6=0,
\eeq
where
\begin{eqnarray}
\nonumber\mathfrak{a}^{irq}_{\phantom{irq}p}&=&\tilde\gamma^{qr}\delta^i_p + \frac{1}{6}\left(\tilde\gamma^{ir}\delta^q_p + \tilde\gamma^{iq}\delta^r_p\right)\\
\nonumber\mathfrak{b}^{iq}_{\phantom{iq}p}&=&\gG{i}{q}{j}{j}{p} - \gG{j}{k}{q}{j}{k}\delta^i_p + \gG{j}{q}{i}{j}{p} + \gG{q}{k}{i}{k}{p} - \frac{2}{3}\gG{i}{q}{k}{k}{p}\\
\mathfrak{c}^{i}_{\phantom{i}p}&=&\tilde\gamma^{ik}\left(\tilde\Gamma^{j}_{kp,j} -\tilde\Gamma^{l}_{jk}\tilde\Gamma^{j}_{lp} + \tilde\Gamma^{j}_{jl}\tilde\Gamma^{l}_{kp}\right)\nonumber\\
&&+\tilde\gamma^{jk}\left(\tilde\Gamma^{i}_{kp,j} -\tilde\Gamma^{l}_{jk}\tilde\Gamma^{i}_{lp} + \tilde\Gamma^{i}_{jl}\tilde\Gamma^{l}_{kp}\right)\\
\nonumber&&-\frac{2}{3}\tilde\gamma^{ij}\left(\tilde\Gamma^{k}_{kp,j} -\tilde\Gamma^{l}_{jk}\tilde\Gamma^{k}_{lp} + \tilde\Gamma^{k}_{jl}\tilde\Gamma^{l}_{kp}\right)\\
\nonumber\mathfrak{d}^i&=&\tilde\nabla_jM^{ij}\\
\nonumber\mathfrak{e}^i&=&-\frac{2}{3\sigma^3}\tilde\gamma^{ij}\partial_j K.
\end{eqnarray}
In order to derive particular numerical values for this set of conformal coefficients, we first need to specify the conformal factor $\Omega_{\rm Kerr}$ that we take for the conformal decomposition of the unperturbed Kerr metric in ACMC-slices. We put
\[\Omega_{\rm Kerr}= \frac{\sigma}{4M}, \]
leading via (\ref{eq:decomposition_Omega}) with $\omega=\sigma$ to $\phi_{\rm Kerr}\equiv 2\sqrt{M}$. Then all coefficients denoted in Fraktur letters as well as $\{H_1,\ldots,H_7\}$ follow explicitly from the Kerr solution in ACMC-slices. In particular, the quantities $\mathfrak{e}^i$ are bounded at \scri by virtue of the ACMC-condition [cf.~equation (\ref{eq:aCMC_K_Taylor})]. The coefficients $\mathfrak{m}^{ij}$ are obtained by requiring that the unperturbed Kerr solution satisfies the momentum constraints through $V^i_{\rm Kerr}\equiv 0$. Using (\ref{eq:decomposition_A}), we conclude that
\[\mathfrak{m}^{ij}=\tilde{A}^{ij}_{\rm Kerr}.\]
Now, for solving the constraints uniquely, we still need to impose boundary conditions. As the momentum constraints form a non-degenerate system of elliptic equations, we may simply choose trivial Dirichlet boundary conditions for the potentials $V^i$,
\begin{align}
\label{eq:Num:Dirichlet_BC}
V^i\Big|_{\scri} =0 = V^i\Big|_{\cal H}.
\end{align}
Note that other choices are possible. Indeed any condition that is compatible with $V^i_{\rm Kerr}\equiv 0$ in the unperturbed case will do\footnote{In this paper we are interested in a single black hole perturbed away from equilibrium, i.e.~from the Kerr solution. In a binary situation, it is frequently desired to impose {\em quasi-equilibrium}, i.e.~non-expanding-horizon and full isolated horizon-like conditions as described in~\cite{York1999,Cook:2000lr, Cook2002, Pfeiffer2003, Pfeiffer:thesis, Cook2004, Jaramillo:2009cc}.}. In contrast, the potential $\phi$ is not subject to freely specifiable boundary conditions. On the one hand, for smooth solutions the Yamabe-like Hamiltonian constraint provides us with a condition valid at \scri,
\beq
\label{eq:Num:Scri_BC}
\phi\Big|_{\scri}=\left[\sqrt[4]{9\tilde\gamma^{11} K^{-2}}\right]_{\scri},
\eeq
which is equivalent to (\ref{eq:taylor_hamilton_constraint}). On the other hand, we require our inner boundary ${\cal H}$ to be a marginally outer trapped surface, which means that (\ref{eq:inner_boundary_conformal}) yields the boundary condition
\beq
\label{eq:Num:Horizon_BC}
\Big[\tilde s^a\partial_a\phi + \mathfrak{h}_1\phi+\mathfrak{h}_2\phi^3+\mathfrak{h}_3\phi^{-3}\Big]_{\cal H}=0
\eeq
where the $\mathfrak{h}_i$ are defined through
\begin{align}
\mathfrak{h}_1&=\frac{1}{4}\left(\tilde{\nabla}_a\tilde{s}^a - \frac{2}{\sigma}\tilde{s}^1\right)\nonumber\\
\mathfrak{h}_2&=\frac{K}{6\sigma}\\
\mathfrak{h}_3&=-\frac{\sigma^2}{4}\tilde{s}^a\tilde{s}^b\tilde{A}_{ab}.\nonumber
\end{align}
As already mentioned, we impose this condition at $\sigma=\sigma_{\cal H}\ne\sigma_{\rm h}$ [see Eq.~(\ref{eq:sh})], in order to obtain {\em perturbed} Kerr data. This is the only perturbation we perform, though various other choices are possible.

We solve the coupled elliptic system (\ref{eq:Num:Hamilt_Constraint}, \ref{eq:Num:Mom_Constraint}) together with the boundary conditions (\ref{eq:Num:Dirichlet_BC}, \ref{eq:Num:Scri_BC}, \ref{eq:Num:Horizon_BC}) by means of a single-domain pseudo-spectral Gauss-Lobatto method. For prescribed numerical resolutions $n_\sigma$ and $n_\mu$, the collocation points are given through
\begin{align}
\sigma_j=&\sigma_{\cal H}\sin^2\left(\dfrac{\pi j}{2N_\sigma}\right)\quad&(j=0,\ldots, N_\sigma)\nonumber\\
  & & \label{eq:Gauss_Lobatto}\\
\mu_k=&-\cos\left(\dfrac{\pi k}{N_\mu}\right) &(k=0,\ldots, N_\mu)\nonumber\,,
\end{align}
where $N_\sigma=n_\sigma-1$ and $N_\mu=n_\mu-1$. The collocation points $\sigma_j$ and $\mu_k$ are the abscissa values of the local extrema of the Chebyshev polynomials $T_{N_\sigma}(2\sigma/\sigma_{\cal H}-1)$ and $T_{N_\mu}(\mu)$ respectively\footnote{As our perturbations lead to equatorially symmetric solutions, the restricted interval $\mu\in[0,1]$ would suffice. However, we have chosen to keep the full interval to allow for non-equatorially symmetric perturbations that might become relevant in the future.}. As described in detail e.g.,~in \cite{meinel2008relativistic}, the values $(\phi_{jk},V^1_{jk},V^2_{jk},V^3_{jk})$  of the unknown potentials at these collocation points are combined to form a vector $\mathbf{f}^{(n_\sigma,n_\mu)}$. From any such vector, the Chebyshev coefficients of the potentials as well as those of their first and second derivatives can be computed. The combination of constraints and boundary conditions, evaluated at the points (\ref{eq:Gauss_Lobatto}), yields a non-linear system of algebraic equations of order $4\times n_\sigma\times n_\mu$ for the entries of the vector $\mathbf{f}^{(n_\sigma,n_\mu)}$. We solve this system by means of a Newton-Raphson scheme, which uses the iterative ``bi-conjugate gradient stabilized method'' \cite{Barrett:1993} for inverting the Jacobian. This method works well only if an appropriate preconditioner is supplied. Here we utilize a finite difference representation of the Jacobian and invert it with the help of a band diagonal matrix decomposition algorithm (see, e.g., \cite{Press2007numericalrecipes} and references therein). An initial guess for the Newton-Raphson scheme is taken from the unperturbed solution, $\{\phi_{\rm Kerr}\equiv 2\sqrt{M},\,\,\, V^i_{\rm Kerr}\equiv 0\}$ for which $\sigma_{\cal H}=\sigma_{\rm h}$. We depart gradually from this solution by modifying the location $\sigma_{\cal H}$ of the inner boundary in several steps. Any solution computed serves as the Newton-Raphson guess for the subsequent solution. In this way we can obtain large perturbations, as will be depicted in the next section. 

\section{Numerical results}
\label{sec:Results}

For any prescribed rotation parameter $j$ with \mbox{$0\le |j|<1$}, the numerical scheme permits the construction of initial data describing perturbations of the corresponding Kerr black hole. The deviation is tuned through the prescription of the location $\sigma_{\cal H}$ of the inner boundary. Values of sample data are given in Table~\ref{tab:examples}:

\begin{table}[h]
\caption{Sample data for the construction of the initial data.}
\label{tab:examples}
\begin{center}
\begin{tabular}{|c||c|c|c|}
\hline
\mbox{Calculation No.}&$j$ & $\sigma_{\rm h}$ & $\sigma_{\cal H}$ \\ 
\hline
1&0.5 & 1.0718 & 0.5 \\ 
\hline
2&0.9 & 1.39286 & 1.4 \\ 
\hline
\end{tabular} 
\end{center}
\end{table}

We observe that the sign of the deviation from the unperturbed solution depends on the sign of $(\sigma_{\cal H}-\sigma_{\rm h})$, i.e.,~the location of the inner boundary with respect to the unperturbed solution.

\begin{figure}[b!]
\begin{center}
\includegraphics[clip]{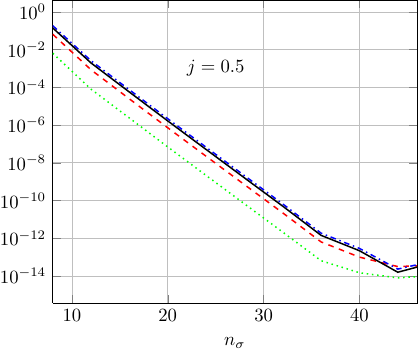}
\includegraphics[clip]{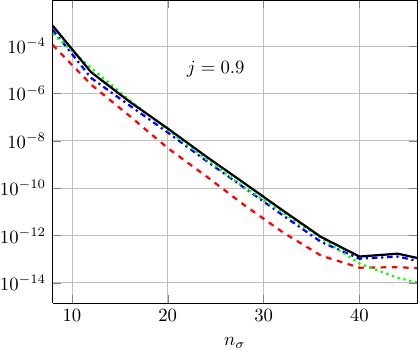}
\caption{For the examples listed in Table \ref{tab:examples}, the maximal deviations $D^\phi_{n_\sigma,n_\mu}$ (red, dashed), $D^1_{n_\sigma,n_\mu}$ (blue, dot-dashed), $D^2_{n_\sigma,n_\mu}$ (green, dotted) and $D^3_{n_\sigma,n_\mu}$ (black, solid) are displayed, see eqns.~(\ref{eqn:D_phi}, \ref{eqn:D_i}). The numerical solutions have been calculated with the resolutions $(n_\sigma,n_\mu)=(n_\sigma,\frac{1}{2}n_\sigma)$.}
\label{fig:supremum_norm}
\end{center}
\end{figure}

In all calculations we find a rapid exponential decay of the Chebyshev coefficients of our numerical solution up to machine precision. Moreover, we see an exponential convergence rate with respect to the numerical resolution 
$(n_\sigma,n_\mu)$. In particular, we monitor the quantities
\begin{eqnarray}
\label{eqn:D_phi} D^\phi_{n_\sigma,n_\mu}&=&M^{-1/2}\sup_{\sigma,\mu}|\phi_{n_\sigma,n_\mu}-\phi_{60,30}|\\
\label{eqn:D_i}  D^i_{n_\sigma,n_\mu}&=&M^{-2}\sup_{\sigma,\mu}|V^i_{n_\sigma,n_\mu}-V^i_{60,30}| 
\end{eqnarray}
which describe the maximal deviation from the numerical reference solution with $(n_\sigma,n_\mu)=(60,30)$. For approximating the supremum, we take coordinate values $\sigma\in[0,\sigma_{\cal H}],\,\,\mu\in[-1,1]$ located on an equidistant grid of 500 times 500 points.
For the examples listed in the table, we provide in figure \ref{fig:supremum_norm} logarithmic plots displaying the deviations $\{D^\phi_{n_\sigma,n_\mu}, D^i_{n_\sigma,n_\mu}\}$. The exponential decay up to round-off error is a strong indication of analyticity of the solutions, i.e.,~for the validity of the criterion (\ref{eq:extrinsic_curvature_scri}). An occurrence of logarithmic terms would have become apparent by a mere power law decay.

\begin{figure}[b!]
\centering
\includegraphics[width=0.49\textwidth,clip]{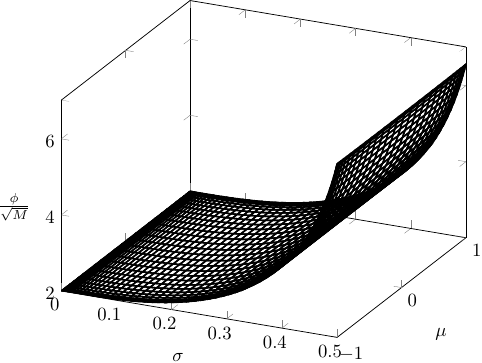}
\includegraphics[width=0.49\textwidth,clip]{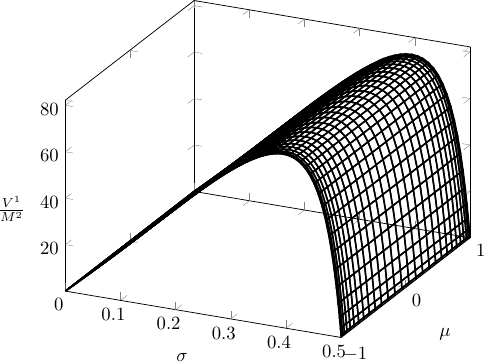}
\includegraphics[width=0.49\textwidth,clip]{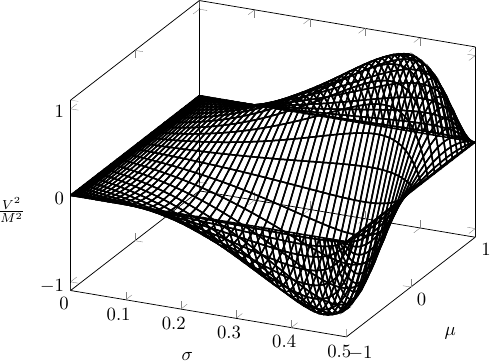}
\includegraphics[width=0.49\textwidth,clip]{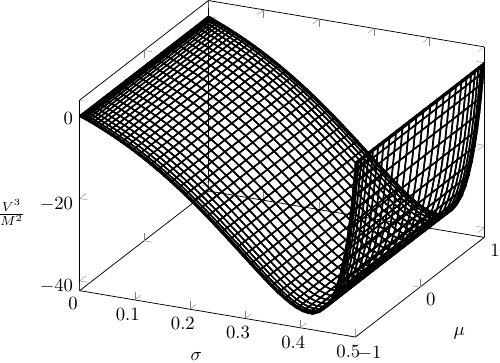}
\caption{The potentials $\{\phi, V^i\}$ for strongly perturbed Kerr initial data are displayed, see example No.~1 in (\ref{tab:examples}). The rotation parameter is $j=0.5$, and the inner boundary is located at $\sigma_{\cal H}=0.5$ which corresponds to $r_{\cal H}\approx 2.14\, \rh$.}
\label{fig:potentials_example}
\end{figure}

Note that the perturbation for calculation no.~1 is large, as $r_{\cal H}=2M/\sigma_{\cal H}\approx 2.14\,\rh$. The corresponding potentials  ($\phi,V^1,V^2,V^3$) are displayed in figure \ref{fig:potentials_example}. In contrast, the perturbation is small for the example no.~2. 

In general, we can assure that our inner boundary is an apparent horizon if we choose the perturbation to be small enough. However, for large perturbations, the outermost marginally outer trapped surface may be located further out. In order to identify the apparent horizon in these cases, we introduce below a horizon finder and perform a careful analysis of our solutions. We verify whether the constructed data correspond indeed to {\em perturbed} Kerr black holes and then access some important physical quantities of the slice, such as the Bondi Mass and the intrinsic angular momentum of the perturbed black hole.

\section{Geometric and physical properties}
\label{sec:Physical}

\subsection{Marginally trapped surfaces}\label{sec:HorizonFinder}
As explained in section \ref{sec:MOTS}, we require the inner numerical boundary  ${\cal H}$ to be a marginally outer trapped surface given by $\Theta_+ = 0.$ Now, it follows directly from the conditions provided by the Hamiltonian constraint when considered at \scri, that the similar condition~$\Theta_-\Big|_{\scri} = 0$ is fulfilled. For small perturbations around the Kerr solution, one can assure that ${\cal H}$ is still the apparent horizon, i.e., it  describes the {\em outermost} marginally outer trapped surface and that $\scri$ is the only surface in our numerical domain satisfying $\Theta_-=0$. However, this picture changes as we allow strong perturbations, since they lead to the formation of new closed 2-surfaces ${\cal S}_+$ and ${\cal S}_-$ inside the numerical grid, satisfying $\Theta_+=0$ and $\Theta_-=0$ respectively.

The marginally trapped surface condition, given by $\Theta_+ \Theta_- = 0$~\cite{Jaramillo:2011,Penrose65}, is satisfied when either $\Theta_+$ or $\Theta_-$ vanishes.  To locate both kinds of 2-surfaces, ${\cal S}_+$ and ${\cal S}_-$, we implemented a horizon finder (see e.g. \cite{Thornburg2005} for a review). Let ${\cal S}_\pm$ be given by 
\[{\cal S}_\pm=\{(\sigma,\mu,\varphi)\in\Sigma\mid \sigma=H_\pm(\mu), 
		\,\,\mu\in[-1,1],\varphi\in[0,2\pi)\}\]
where $\Sigma$ denotes the initial hyperboloidal ACMC-slice described by the coordinates $(\sigma,\mu,\varphi)$. Writing the outward pointing normal-vector of this surface as
\beq\nonumber
\tilde{s}^i =-\tilde B\tilde\nabla^i\left[\sigma-H_\pm(\mu)\right]=-\tilde B(\tilde\gamma^{1i}-H_\pm'\tilde\gamma^{2i}),\label{eq:MTS_s}
\eeq
where $H_\pm'=dH_\pm/d\mu$ and $\tilde B$ follows from the normalization condition $\tilde s^i\tilde s_i=1$,
\begin{equation}\nonumber
\tilde{B}^{-2}=\tilde\gamma^{11} -2H_\pm'\tilde\gamma^{12} + (H_\pm')^2\tilde\gamma^{22},\label{eq:MTS_norm}
\end{equation}
the trapped surface condition (\ref{eq:Expansions}) turns into an ordinary differential equation for $H_\pm$. Again, we solve this equation with a pseudo-spectral method as the one presented at the end of section \ref{sec:Numeric}. 

\begin{figure}[b!]
\begin{center}
\includegraphics[width=0.33\textwidth,clip]{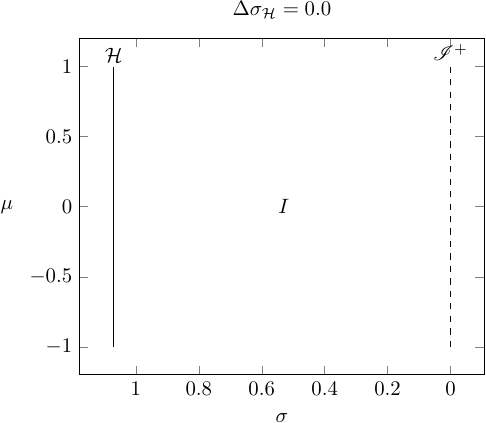}
\includegraphics[width=0.31\textwidth,clip]{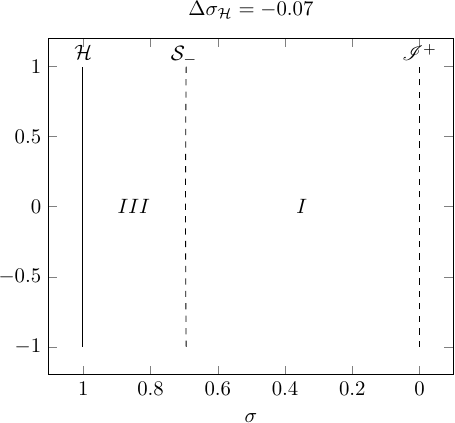}
\includegraphics[width=0.31\textwidth,clip]{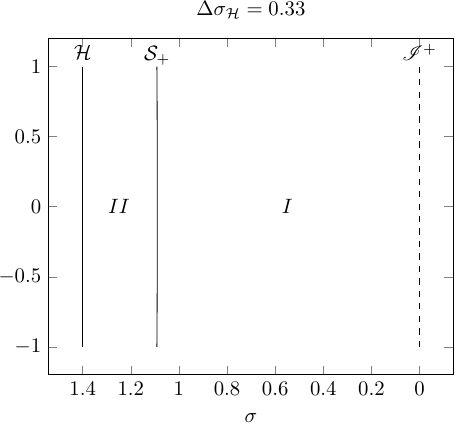}
\caption{Marginally trapped surfaces $\Theta_+=0$ (solid line) and $\Theta_-=0$ (dashed line). The left panel presents the expected behaviour for the unperturbed Kerr spacetime with $j=0.5$, where the only solutions obtained by the horizon finder are the inner ${\cal H}$ and outer $\scri$ numerical boundaries. Small perturbations ($\Delta\sigma_{\cal H}  \gtrsim 0$ or $\Delta\sigma_{\cal H}  \lesssim 0$) around the Kerr space-time have the same features. Strong perturbations in the negative direction ($\Delta\sigma_{\cal H}<0$) lead to the formation of another solution ${\cal S}_-$  for $\Theta_-=0$, apart from $\scri$, while strong perturbations in the positive direction ($\Delta\sigma_{\cal H} > 0$) lead to the formation of a new solution ${\cal S}_+$  for $\Theta_+=0$, other than ${\cal H}$. The $\mu$-dependence is very mild (see Fig.~\ref{fig:MTS_zoom}). The regions $I,II$ and $III$ are explained in the text.
}
\label{fig:MTS}
\end{center}
\end{figure}

\begin{figure}[b!]
\begin{center}
\includegraphics[width=7.8cm,clip=true]{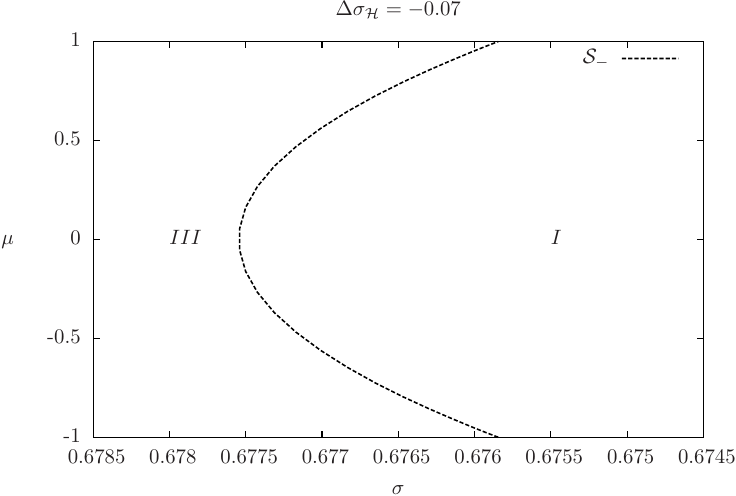}
\includegraphics[width=7.6cm,clip=true]{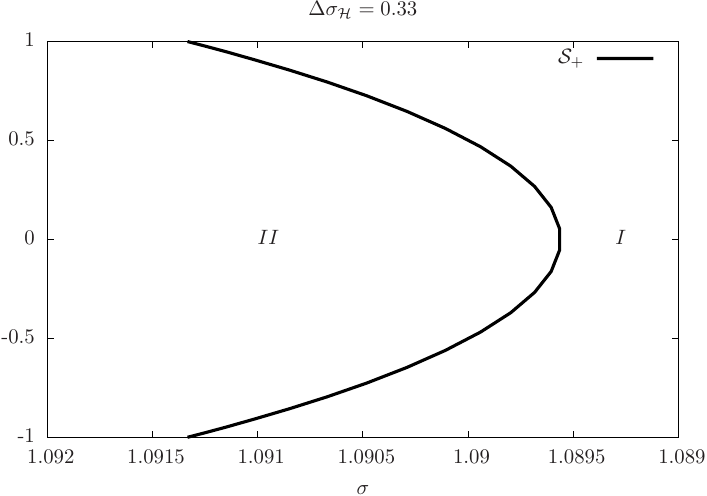}
\caption{Shown is the $\mu$-dependence of the functions $H_-$ and $H_+$ describing the trapped surfaces ${\cal S_-}$ (dashed line in the left panel) and ${\cal S_+}$ (solid line in the right panel). The middle and right panel of Fig.~\ref{fig:MTS} are displayed here in a narrower range of the coordinate $\sigma$.
}
\label{fig:MTS_zoom}
\end{center}
\end{figure}

Fig.~\ref{fig:MTS} summarizes the results of the horizon finder in terms of the perturbation parameter $\Delta\sigma_{\cal H} = \sigma_{\cal H} - \sigma_{\rm h}$. The left panel shows the structure of the marginally trapped surfaces for the unperturbed Kerr solution with $j=0.5$. As expected, apart from the inner ${\cal H}$ and outer $\scri$ numerical boundaries, no other surfaces with $\Theta_+=0$ or $\Theta_-=0$ were found. For small perturbations ($\Delta\sigma_{\cal H}  \gtrsim 0$ or $\Delta\sigma_{\cal H}  \lesssim 0$), this picture does not change and the numerical boundaries coincide with the solution of the horizon finder. 

In contrast, the middle panel of Fig.~\ref{fig:MTS} presents a strongly perturbed black hole with the new inner boundary  $\cal H$ at a smaller coordinate value $\sigma_{\cal H} < \sigma_{\rm h}$ ($\Delta\sigma_{\cal H}<0$), corresponding to a larger coordinate radius $r_{\cal H}=1/\sigma_{\cal H}$. While $\cal H$ is still the only surface satisfying $\Theta_+=0$, the horizon finder locates, apart from $\scri$, a second surface ${\cal S}_-$. Similarly, as shown in the right panel of Fig.~\ref{fig:MTS}, strongly perturbed black holes with $\sigma_{\cal H} > \sigma_{\rm h}$ ($\Delta\sigma_{\cal H}>0$) possess $\scri$ as the only surface with $\Theta_-=0$ but, apart from ${\cal H}$, another surface ${\cal S}_+$. Note that the $\mu$-dependence of $H_\pm$ is hardly recognized in Fig.~\ref{fig:MTS}. However, by restricting ourselves to a narrower range of the coordinate $\sigma$ (see Fig.~\ref{fig:MTS_zoom}), one can clearly see that both $H_\pm$ depend very mildly on $\mu$.

For a fixed free-data parameter $j$, one can start from the unperturbed case $\Delta\sigma_{\cal H}=0$ and slowly increase the perturbation parameter $\Delta\sigma_{\cal H}$. At a critical value $\Delta\sigma_{\cal H}^{II}$, ${\cal S}_+$ coincides exactly with the numerical boundary ${\cal H}$ and for $\Delta\sigma_{\cal H}>\Delta\sigma_{\cal H}^{II}$, ${\cal S}_+$ enters the numerical domain and becomes the apparent horizon.

A similar behavior is observed for ${\cal S}_-$. For small perturbations, ${\cal S}_-$ is not found. Here however, there is no critical parameter for which ${\cal S}_-$ and ${\cal H}$ coincide exactly. Rather, there is a transition phase during which ${\cal S}_-$  and ${\cal H}$ intersect each other. The critical value $\Delta\sigma_{\cal H}^{III}$ is then defined as the one at which ${\cal S}_-$ lies entirely within the numerical domain and touches ${\cal H}$. For $\Delta\sigma_{\cal H}<\Delta\sigma_{\cal H}^{III}$, ${\cal S}_-$ moves inside the numerical domain.

In the initial slice $\Sigma$, the surfaces ${\cal S}_+$ and ${\cal S}_-$ are, respectively, the boundaries of the  trapped regions $II$ and $III$ (see Fig.~\ref{fig:MTS}). This follows from the fact that one can always find closed 2-surfaces inside these regions, along which the corresponding expansions $\Theta_\pm$ have uniform signs obeying the following characterization:
\beq
\begin{array}{llll}
	\mbox {Region $I$:}    &\qquad \Theta_+>0  & \mbox{and} & \Theta_-<0,\\[2mm]
	\mbox {Region $II$:}   &\qquad \Theta_+<0  & \mbox{and} & \Theta_-<0,\\[2mm]
	\mbox {Region $III$:}  & \qquad \Theta_+>0 & \mbox{and} & \Theta_->0.
\end{array}
\eeq
The critical parameters $\Delta\sigma_{\cal H}^{II}$ and $\Delta\sigma_{\cal H}^{III}$, which are associated with the formation of regions $II$ and $III$ inside our numerical domain, depend on the rotation parameter $j$. For example, for $j=0.5$ we obtain $\Delta\sigma_{\cal H}^{II}\approx0.074$ and $\Delta\sigma_{\cal H}^{III}\approx-0.030$. 

Region $II$ is classified as {\em future} trapped region and is associated with the interior of the black hole, since both outgoing and ingoing light rays converge~\cite{Jaramillo:2011}. As we are interested in initial data outside the black hole region, one can restrict the numerical domain up to $S_+$ and consider this surface as the inner numerical boundary for perturbations in the regime $\Delta\sigma_{\cal H}>\Delta\sigma_{\cal H}^{II}$. On the other hand, region $III$ is classified as a {\em past} trapped region. Here, outgoing and ingoing light rays diverge, which corresponds to a local characterization of a white hole. In particular, at the surface ${\cal S}_-$ 
it holds that $\Theta_-=0,\,\Theta_+>0$ and $\delta_+\Theta_-<0$ (meaning that $\Theta_-$ changes from positive to negative values while passing from region $III$ into region $I$, cf.~Fig.~\ref{fig:MTS}). 
Following the quasi-local horizon formalism presented in \cite{Hay94}, we conclude that ${\cal S}_-$ is a section of a `past outer trapping horizon', 
indeed the horizon of a white hole. We expect that initial data possessing region $III$ outside the apparent horizon need to be discarded on physical grounds, as there is up to now no known realistic  astrophysical scenario which would lead to the formation of white holes. However, such data could potentially be of interest in the stability analysis of the Kerr solution, as they provide a probe of perturbed white hole initial data which are in principle not accessible
via a physical collapse scenario. In this context, the explicit construction of data containing 
an `apparent horizon' $\Theta_+=0$ surface, namely ${\cal H}$, contained inside an exterior 
$\Theta_-=0$ white hole horizon, namely ${\cal S_-}$, could be of interest in the study of the
stability of the bifurcation surface in the Kerr solution separating the black hole from the white hole
regions.

\subsection{Mass and angular multipole moments}\label{sec:Moments}

Having obtained the numerical solution, we perform a source multipolar decomposition of the apparent horizon ${\cal H}$\footnote{As observed in the previous section, for small perturbation, the apparent horizon coincides with the numerical boundary ${\cal H}$, while strong perturbations lead to the formation of a new trapped surface ${\cal S}$. In this section, we restrict ourselves to the small perturbation regime where ${\cal H}$ still describes the apparent horizon.}. Introduced in~\cite{Ashtekar:2004gp}, the multipole moments of isolated horizons are an invariant measure on ${\cal H}$ that allows us to tell if the new initial data represent, indeed, a deformed black hole or if they are still associated with the (unperturbed) Kerr metric (with possibly new mass $M$ and specific angular momentum $j$) expressed in some different coordinates \cite{Vasset:2009pf}. In fact, as shown in~\cite{Ashtekar:2004gp}, the geometry of a given isolated horizon is completely characterized by multipole moments divided into two types: mass multipoles $M_n$ and angular multipoles $J_n$. In the case of dynamical horizons, they were first employed in~\cite{Schnetter-Krishnan-Beyer-2006} to verify that during the numerical evolution of a space-time, the black hole settles down to a Kerr solution.  

A gauge-independent definition of such multipoles relies on the existence of an axial symmetry $\Phi^i$ on ${\cal H}$, which here is given by the axial Killing vector $(\partial_\varphi)^i$. Another important issue for their construction is the introduction of a new coordinate system $\{\hat{\mu}, \hat{\varphi}\}$ on ${\cal H}$, in such a way that the Legendre polynomials $P_n(\hat{\mu})$ hold the correct orthonormality properties. As described in~\cite{Ashtekar:2004gp, Schnetter-Krishnan-Beyer-2006}, $\hat{\varphi} \in [0, 2\pi)$ is the affine-parameter along $\Phi^i$ which coincides, in our case, with the original coordinate $\varphi$. Apart from that, $\hat{\mu} = \cos{\hat{\theta}} \in [-1,1]$ is defined by
\beq
\label{eq:newmu}
\partial_i \hat{\mu} = \frac{1}{R_{{\cal H}}^2}{}^{(2)}\epsilon_{ji}\Phi^j,
\eeq
with ${}^{(2)}\epsilon_{j i}$ being the area-element form on ${\cal H}$ and $R_{{\cal H}}$  the areal radius defined by the area of ${\cal H}$, i.e.  \[{\cal A}_{{\cal H}} = \oint_{{\cal H}} dA = 4\pi R_{{\cal H}}^2.\] The integration constant in (\ref{eq:newmu}) is determined by requiring that \[\oint_{{\cal H}} \hat{\mu}dA = 0.\]
\begin{figure}[h!]
\begin{center}
\includegraphics[width=7.7cm,clip=true]{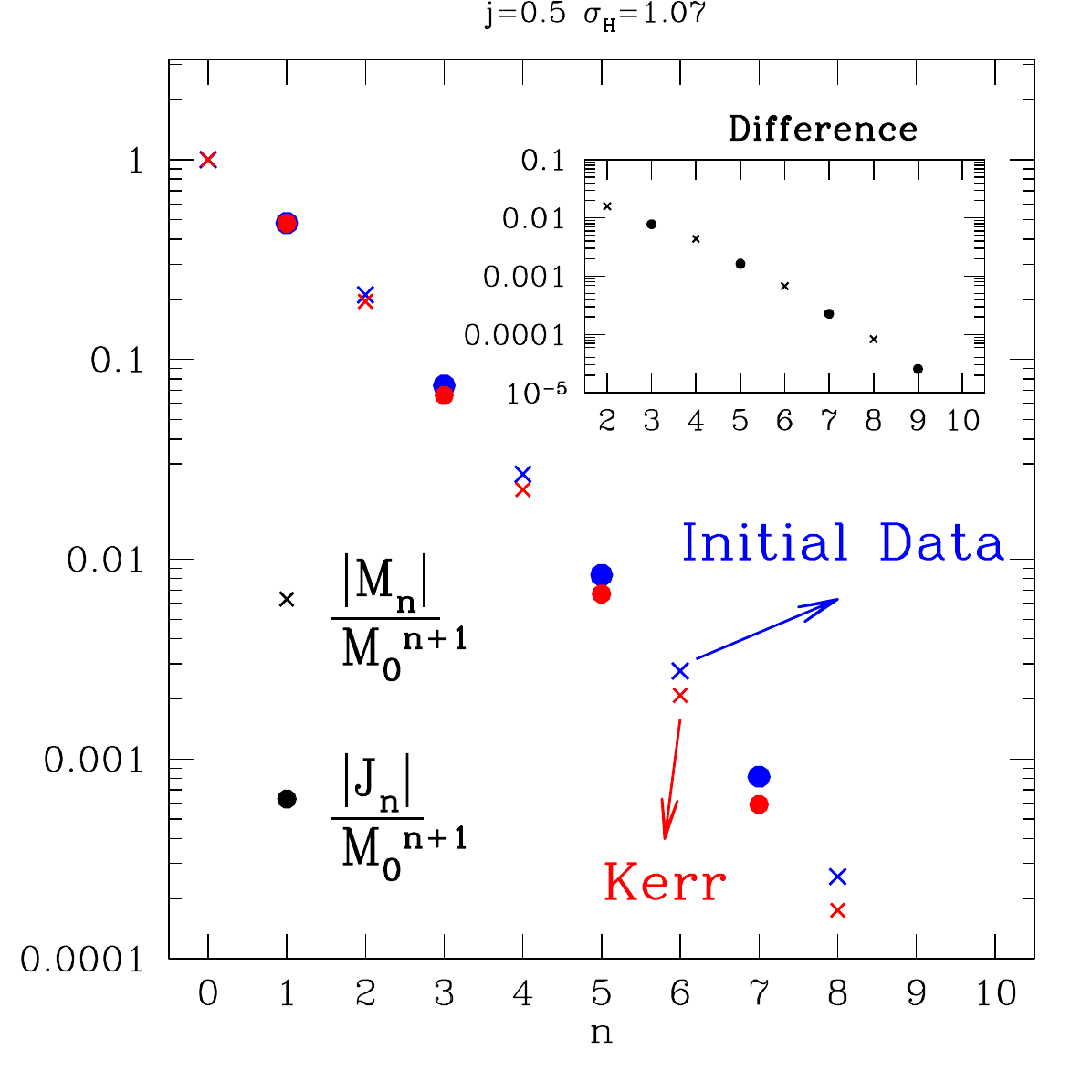}
\includegraphics[width=7.7cm,clip=true]{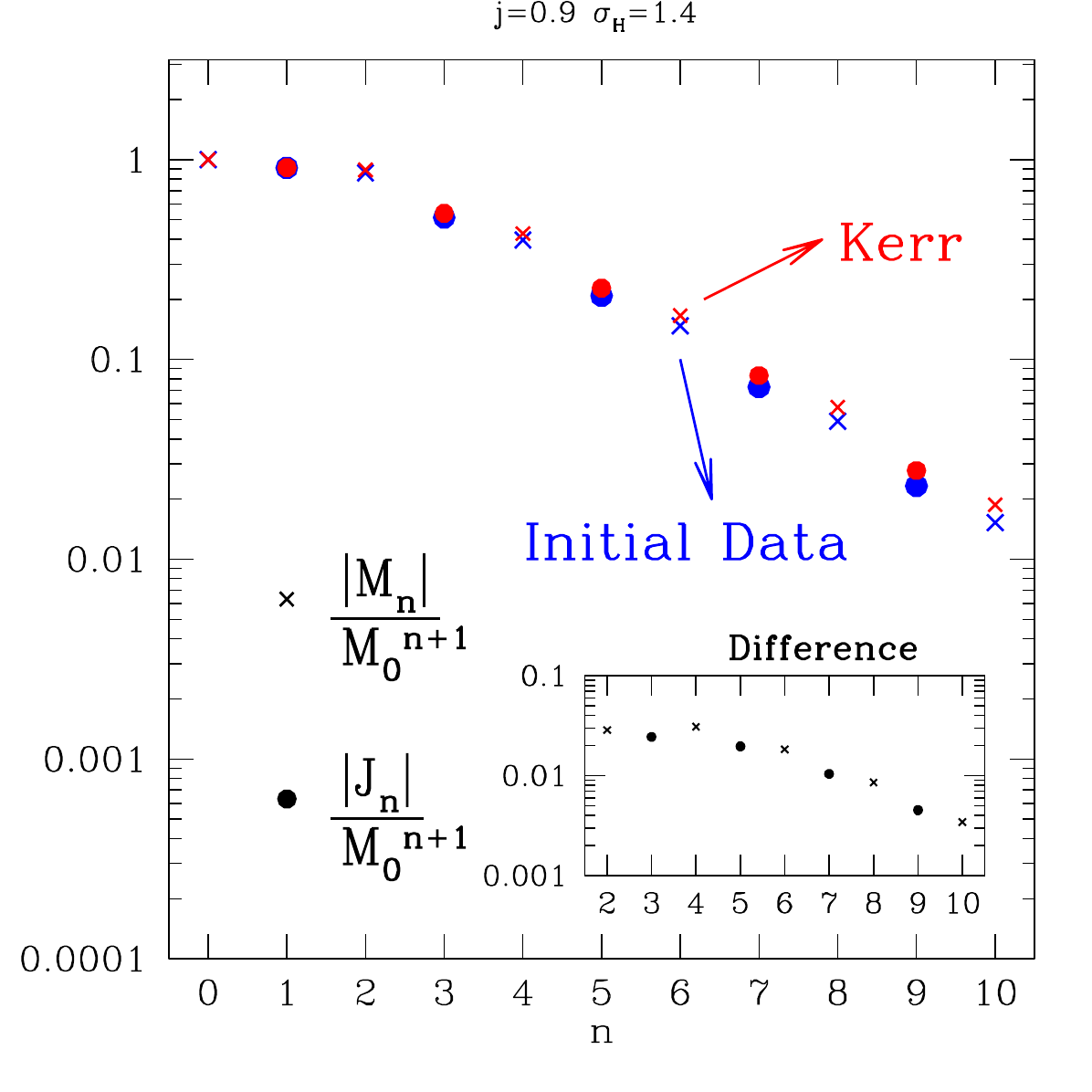}
\includegraphics[width=7.7cm,clip=true]{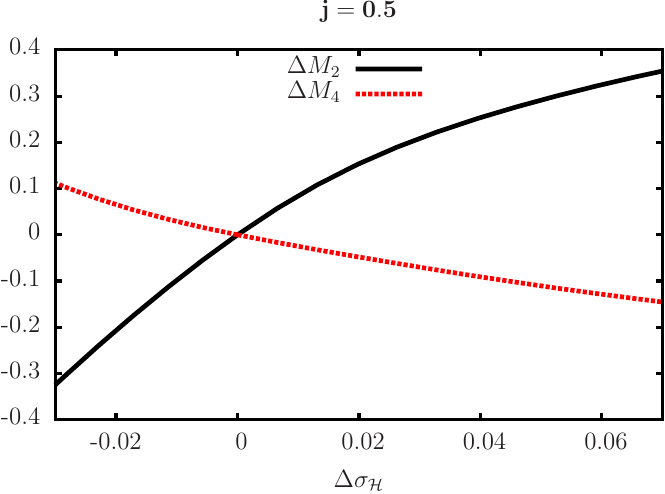}
\includegraphics[width=7.7cm,clip=true]{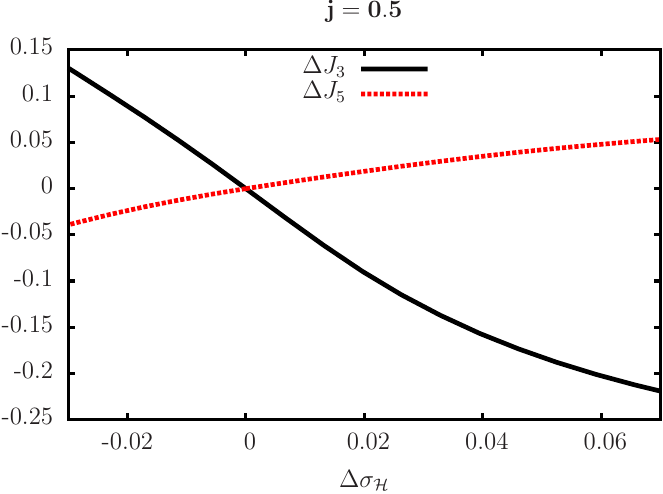}
\end{center}
\caption{At the top, the left and right panels show, respectively, the multipole moments for solutions obtained with default rotation parameters $j=0.5$ and $j=0.9$. Inner boundaries were placed, respectively, at $\sigma_{\cal H}=1.07$ and $\sigma_{\cal H}=1.4$. Crosses and dots are associated, respectively, with the normalized mass $\left|M_n\right|/M_0^{n+1}$ and the angular $\left|J_n\right|/M_0^{n+1}$ moments, while the inset shows the difference between their values for the perturbed black hole and the corresponding Kerr solution with same mass $M_0$ and angular momentum $J_1$. For $j=0.5$, such differences as functions of the perturbation parameter $\Delta \sigma_{\cal H}$ for the mass moments $M_2$ and $M_4$ (left panel) and the angular moments $J_3$ and $J_5$ (right panel) are shown at the bottom. The deviations from the corresponding Kerr values demonstrate that our initial data do not correspond to the Kerr solution expressed in different coordinates. }
\label{fig:MultSmallPert}
\end{figure}
The angular and mass multipoles are then defined as 
\bea
J_n & = & \frac{R_{{\cal H}}^{n-1}}{8\pi}\oint_{{\cal H}} K_{\Phi s}(\mu) P'_n \left(\hat{\mu}(\mu)\right)dA \label{eq:AngMult} \\
M_n & = & M_{{\cal H}}\frac{R_{{\cal H}}^{n}}{8\pi}\oint_{{\cal H}} {}^{(2)}R(\mu)P_n(\hat{\mu}(\mu))dA \label{eq:MassMult}. 
\eea

In Eq.~(\ref{eq:AngMult}), $K_{\Phi s} = K_{ij}\Phi^i s^j$ where
$s^i$ is the unit vector normal to ${\cal H}$. The derivative of $P_n$ should be taken with respect to its argument $\hat{\mu}$. Apart from that, in Eq.~(\ref{eq:MassMult}), ${}^{(2)}R$ is the Ricci scalar of the two-surface ${\cal H}$ and the parameter $M_{{\cal H}}$ is given by
$M_{{\cal H}} = \sqrt{R^2_{{\cal H}}/4 + J_1^2/R_{{\cal H}}^2}$, which corresponds to the black hole mass in case of the Kerr space-time.

Using the Gauss-Bonnet theorem, it is not hard to convince oneself that $M_0=M_{{\cal H}}$. One can also show that $M_1=0$~\cite{Ashtekar:2004gp,Jaramillo:2011re}. Regarding the angular multipoles, $J_0$ is clearly zero and $J_1$ is exactly the definition of the Komar angular momentum on the horizon, which will be used in the next section to describe the black hole's angular momentum, $J\equiv J_1$. Finally, due to the symmetry across the equatorial plane of our initial data, we have that for all $m\in\mathbb{N}$ the multipoles $M_{2m+1}$ and $J_{2m}$ vanish.

Given the numerical initial data, calculated as described in section \ref{sec:Results}, we evaluate the multipole moments (\ref{eq:AngMult}) and (\ref{eq:MassMult}) with respect to the inner numerical boundary ${\cal H}$. We restrict ourselves to small perturbations, in order to assure that ${\cal H}$ still describes the apparent horizon. We compare our initial data with a corresponding Kerr space-time possessing the same mass $M_0$ and angular momentum $J_1$. Then, the distinctness of higher multipole moments allows us to identify a perturbed black hole. Note that the multipoles have  dimensions of ${\rm [Mass]}^{n+1}$, so we present our results in terms of the normalized quantities $J_n/M_0^{n+1}$ and $M_n/M_0^{n+1}$.

Fig.~\ref{fig:MultSmallPert} depicts results for small perturbations around the Kerr solution. In particular, in the upper-left and upper-right panels, results are displayed which correspond to the specified rotation parameters $j=0.5$ and $j=0.9$ respectively. The  coordinate locations of the horizon were placed at $\sigma_{\cal H}=1.07$ and $\sigma_{\cal H}=1.4$. The figure also shows the differences $\Delta M_n$ and $\Delta J_n$ between the moments for the perturbed black hole and the corresponding Kerr solution with same mass $M_0$ and angular momentum $J_1$. While the insets in the top panels show these differences for a fixed coordinate location of the new apparent horizon, the bottom panels depict $\Delta M_2$ and $\Delta M_4$ (left panel) and $\Delta J_3$ and $\Delta J_5$ (right panel) as functions of the perturbation parameter $\Delta \sigma_{\cal H}$ for $j=0.5$. We find that the multipole moments confirm our expectation that the new initial data do indeed represent a deformed rotating Kerr black hole. 

\subsection{Bondi Mass}\label{sec:BondiMass}

We now calculate the energy content of our slice by evaluating the Bondi Mass~\cite{Bondi1962}, as given by the value of the Hawking Mass at $\scri$~\cite{Szabados04,Jaramillo:2010ay,Rinne:2013qc}.  For a given closed 2-surface $S$, the Hawing Mass is defined as 
\beq
\label{eq:HawkMass}
M_H= \sqrt{\frac{{\cal A}_{S}}{16\pi}} \left[ 1 + \frac{1}{8\pi}\oint_{S}    \Theta_{+} \Theta_{-} dA   \right],
\eeq
where ${\cal A}_{ S}$ is the area of $S$ (similar to ${\cal A}_{\cal H}$ in the previous section). Also, $\Theta_{+} $ and $\Theta_{-} $
are, respectively, the expansions of outgoing and ingoing null vectors.

Considering coordinate surfaces characterized by constant $\sigma$-values, eq.~(\ref{eq:HawkMass}) can be re-written in terms of conformal quantities as
\beq
\label{eq:HawkMass_conf}
M_H= \sqrt{\frac{{\cal \tilde{A}}_S}{16\pi}}\Omega^{-1} \left[ 1 + \frac{1}{4\pi}\oint_{S}    I \d \tilde{A}   \right],
\eeq
with the integrand
\bea
I & = & \sigma^{-2}\left[\frac{K^2}{9}\phi^4 - (\tilde{\gamma}^{11})^2 \right] - \sigma^{-1}\sqrt{\tilde{\gamma}^{11}}\left[ \tilde{\kappa}-4\frac{\tilde{\gamma}^{1i}}{\sqrt{\tilde\gamma^{11}}}\frac{\phi_{,i}}{\phi} \right] \nonumber \\
& - & \frac{1}{4}\left[ \tilde{\kappa}-4\frac{\tilde{\gamma}^{1i}}{\sqrt{\tilde\gamma^{11}}}\frac{\phi_{,i}}{\phi} \right]^2 + \frac{K\tilde{q}^{ij}\tilde{A}_{ij}}{3\phi^2}\sigma + \frac{\left(\tilde{q}^{ij}\tilde{A}_{ij}\right)^2}{4\phi^4}\sigma^4.
\eea

As in section \ref{subsec:RegCond} (see equations (\ref{eq:kappa}) and (\ref{eq:q_ij})), we introduced above the induced conformal metric $\tilde q_{ij}$ and corresponding mean curvature $\tilde\kappa=\tilde q^{ij}\tilde\kappa_{ij}$ where $\tilde\kappa_{ij}$ is the induced extrinsic curvature, defined at the closed 2-surface $S$.

Taking into account the asymptotic conditions of the constraint equations discussed in section \ref{subsec:RegCond}, one can show that the terms proportional to $\sigma^{-2}$ and $\sigma^{-1}$ assume finite values at \scri. The Bondi Mass of our ACMC slice can then be written as $\displaystyle M_{\rm B} = \lim_{\sigma\rightarrow 0}M_H$, and it follows that
\beq
\label{eq:BondiMass}
 \frac{M_B}{M}=1+ \frac{1}{2}\int_{-1}^1  \delta I_1 \sqrt{\tilde q} \d \mu
\eeq
with $\tilde q=\det(\tilde q_{ij})$. The integrand is given in terms of the potentials $(\phi, V^i)$ for the perturbed data,
\[
\delta I_1 = \left[ \frac{\sqrt{\tilde\gamma^{11}} \tilde{q}^{ij}({\cal L}V)_{ij}} {16M^2}    +  \tilde\gamma^{11}\frac{4}{3}\frac{\partial^3_{\sigma}\phi}{\sqrt{M}} \right]_{\scri}.
 \]
 
\begin{figure}[t!]
\begin{center}
\includegraphics[clip]{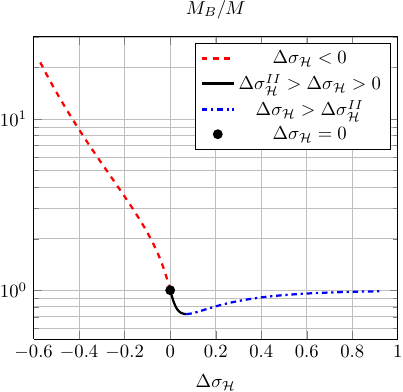}
\end{center}
\caption{$M_{\rm B}/M$ as a function of the perturbation parameter $\Delta\sigma_{\cal H}$. For the unperturbed case (black dot) we have $M_{\rm B} = M$, while $M_{\rm B}>M$ holds if $\Delta\sigma_{\cal H}<0$ (dashed red line). The solid black line corresponds to the regime $0<\Delta\sigma_{\cal H}<\Delta\sigma_{\cal H}^{II}$ where the inner numerical boundary is the apparent horizon. The dash-dotted blue line corresponds to the regime $\Delta\sigma_{\cal H}>\Delta\sigma_{\cal H}^{II}$ for which the apparent horizon is located inside the numerical domain (cf.~section \ref{sec:HorizonFinder}). The relation $M_{\rm B}<M$ holds for all $\Delta\sigma_{\cal H}>0$, but for strong perturbations the Bondi Mass approaches the scaling parameter $M$.}
\label{fig:BondiMass}
\end{figure}

\begin{figure}[t!]
\begin{center}
\includegraphics[clip]{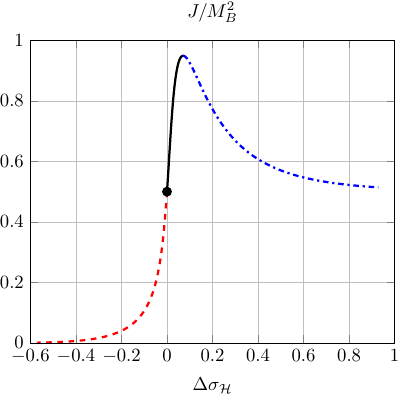}
\includegraphics[clip]{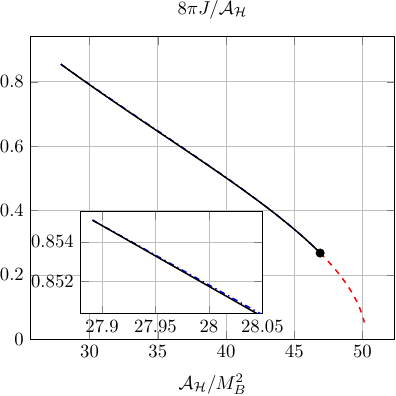}
\end{center}
\caption{Rotation parameters are plotted against the perturbation parameter $\Delta\sigma_{\cal H} = \sigma_{\cal H} - \sigma_{\rm h}$ (left panel) and the coordinate-independent horizon area parameter ${\cal A}_{\cal H}/M_{\rm B}^2$ (right panel). The free data used in the construction are characterized by the default rotation parameter $j=0.5$. Note that for $\Delta\sigma_{\cal H}\sim 0.08$, $j_{\rm B}= J /M_{\rm B}^2$ reaches a maximal value of about $0.95$. Perturbations with $\Delta\sigma_{\cal H}>0$ correspond to larger rotation parameters $j_{\rm B}>j$.
}
\label{fig:MassAngMom}
\end{figure}

 Fig.~\ref{fig:BondiMass} shows the behavior of $M_{\rm B}/M$ as a function of the perturbation parameter $\Delta\sigma_{\cal H}$. Here, we consider $M$ as a scaling parameter in the freely-specifiable quantities which have been taken at the outset from the (unperturbed) Kerr metric. Note that the plot is divided according to the discussion in section \ref{sec:HorizonFinder}. The black dot represents the unperturbed case, which gives $M_{\rm B} = M$. The branch with $\Delta\sigma_{\cal H}<0$ is depicted with a red dashed line, and provides always $M_{\rm B}>M$, irrespective of the formation of the new region $III$ inside the numerical grid. For $\Delta\sigma_{\cal H}>0$, one obtains $M_{\rm B}<M$. The solid black line indicates the regime $0<\Delta\sigma_{\cal H}<\Delta\sigma_{\cal H}^{II}$ where the prescribed inner numerical boundary is the apparent horizon, whereas the dash-dotted blue line describes the results for $\Delta\sigma_{\cal H}>\Delta\sigma_{\cal H}^{II}$, for which the apparent horizon is located inside the numerical grid. Note that the color/typographic code in Fig.~\ref{fig:BondiMass} is also used in the upcoming Figs.~\ref{fig:MassAngMom} and \ref{fig:AreaIneq}. The results correspond to initial data constructed with prescribed rotation parameter $j=0.5$.
 
 Apart from the Bondi Mass, the black hole angular momentum $J$ (given by the moment $J_1$ from the previous section) allows us to also introduce the specific angular momentum of the perturbed black hole via $j_{\rm B} = J_1/M_{\rm B}^2$, which is shown in Fig.~\ref{fig:MassAngMom} in terms of the perturbation parameter $\Delta\sigma_{\cal H}$. Here, one also notes two distinct behaviors depending on the sign of $\Delta\sigma_{\cal H}$. Taking $\Delta\sigma_{\cal H}>0$, one is led to a perturbed black hole with higher spin parameter $j_{\rm B}$ which tends asymptotically to the prescribed parameter value $j=0.5$, while the case $\Delta\sigma_{\cal H}<0$ corresponds to smaller specific angular momentum, $j_{\rm B}<j$. 
 
In the right panel of Fig.~\ref{fig:MassAngMom}, the quantity $8\pi J/{\cal A}_{{\cal H}}$ is plotted against ${\cal A}_{{\cal H}}/M_{\rm B}^2$, thereby avoiding the coordinate-dependent parameter $\Delta\sigma_{\cal H}$.

One can clearly see the two distinct branches which are associated with the situations described above. For $\Delta\sigma_{\cal H}<0$ the horizon area parameter ${\cal A}_H/M_B^2$ increases while the rotation parameters $j_{\rm B}$ and $J/{\cal A}_H$ decrease. Similarly, for sufficiently small values $\Delta\sigma_{\cal H}>0$ the opposite effects occur. However, as $\Delta\sigma_{\cal H}$ is increased further, a maximal value of $j_{\rm B}$ is encountered, followed by a growth of the horizon area parameter and a fall-off of the rotation parameter $j_{\rm B}$. Note that in the right panel of Fig.~\ref{fig:MassAngMom}, the curves corresponding to the two regimes $0<\Delta\sigma_{\cal H}<\Delta\sigma_{\cal H}^{II}$ (solid black line) and $\Delta\sigma_{\cal H}>\Delta\sigma_{\cal H}^{II}$ (dash-dotted blue line) lie almost on top of each other.
 
Interestingly, for large perturbations in the regime $\Delta\sigma_{\cal H}>\Delta\sigma_{\cal H}^{II}$, the Bondi Mass seems to approach asymptotically the scaling parameter $M$. Likewise, also the other quantities $J/M_{\rm B}^2, J/{\cal A}_{{\cal H}}$ and ${\cal A}_{{\cal H}}/M_{\rm B}^2$ come back again to their values in the unperturbed situation. A similar behavior was found in \cite{JarVasAns08} where also deformed Kerr initial data were constructed which seem to return to the unperturbed solution in the limiting case corresponding
to large values of the perturbation parameter. In the future, we intend to investigate this observation further.
  
 \subsection{black hole inequalities}
 
We finish the study of our new initial data with an investigation of some black hole inequalities. Here the aim is twofold. At first, it adds to the discussion in section \ref{sec:Moments} and proves independently that the data correspond to perturbed Kerr black holes. Secondly, it probes some aspects in the standard picture of
gravitational collapse \cite{Jaramillo:2011}. 

Assuming the validity of weak cosmic censorship, Penrose conjectured the inequality relating the black hole area $A_{{\cal H}}$ to the ADM mass~\cite{Penrose73}:
\beq
\label{eq:IneqADMMass}
{\cal A}_{{\cal H}} \leq 16 \pi M_{\rm ADM}^2.
\eeq
Interestingly, the reasoning leading to Penrose's inequality passes through an intermediate stage, 
\beq
\label{eq:IneqBondiMass}
{\cal A}_{{\cal H}} \leq 16 \pi M_{\rm B}^2,
\eeq
which is stronger than (\ref{eq:IneqADMMass}) since the Bondi Mass $M_{\rm B}$ is never larger than the ADM Mass $M_{\rm ADM}$. 

\begin{figure}[t!]
\begin{center}
\includegraphics[clip]{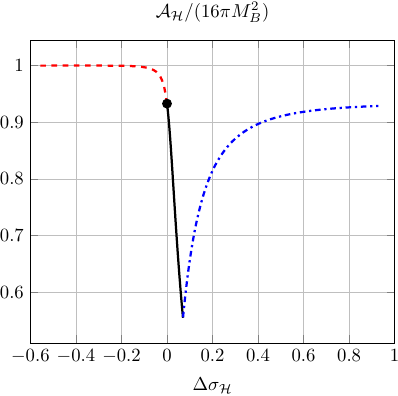}
\end{center}
\caption{Probing the Penrose-Bondi conjecture ${\cal A}_{{\cal H}} \leq 16 \pi M_{\rm B}^2$ for our initial data with default rotation parameter $j=0.5$.
}
\label{fig:AreaIneq}
\end{figure}

While we have no access to the ADM Mass, we can probe inequality (\ref{eq:IneqBondiMass}) with the Bondi Mass calculated above. Fig.~\ref{fig:AreaIneq} shows the behavior of ${\cal A}_{{\cal H}}/(16 \pi M_{\rm B}^2)$ in terms of $\Delta\sigma_{\cal H}$ and demonstrates the validity of (\ref{eq:IneqBondiMass}) for our initial data with prescribed rotation parameter $j=0.5$.

Penrose's conjecture also has a rigidity part, according to which equality in expressions (\ref{eq:IneqBondiMass}) and (\ref{eq:IneqADMMass}) is obtained only for the Schwarzschild solutions. Repeating Penrose's argument in the axisymmetric case, inequality (\ref{eq:IneqADMMass}) can be refined to
\beq
\label{eq:IneqDain}
\epsilon_{A}:=\frac{{\cal A}_{{\cal H}}}{8\pi (M_{\rm ADM}^2 + \sqrt{M_{\rm ADM}^4 - J^2})}\le 1,
\eeq
introduced in Dain et.~al~\cite{Dain2002}. The rigidity conjecture in this case
states that $\epsilon_A = 1$ holds true only for spacetimes in the Kerr
family. Results in favor of the validity of (\ref{eq:IneqDain}) were presented in \cite{JarVasAns08}. Now, utilizing the same arguments as for the Penrose inequality, we substitute the ADM Mass by the Bondi Mass in inequality (\ref{eq:IneqDain}) and propose the definition of a {\em Dain-Bondi number} as
\beq
\epsilon_{AB}=\frac{{\cal A}_{{\cal H}}}{8\pi (M_{\rm B}^2 + \sqrt{M_{\rm B}^4 - J^2})}.\label{eq:DainBondiNumber}
\eeq

 \begin{figure}[t!]
\begin{center}
\includegraphics[clip]{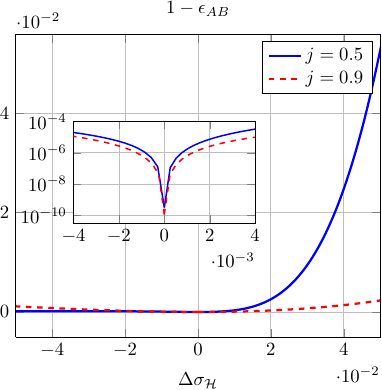}
\end{center}
\caption{Probing the Dain-Bondi inequality $\epsilon_{AB}\leq 1$ with $\epsilon_{AB}$ defined in eq. (\ref{eq:DainBondiNumber}) for initial data sets with default rotation parameters $j=0.5$ and $j=0.9$. Equality (within numerical accuracy) only holds for the unperturbed solution $\Delta\sigma_{\cal H} = 0$.
}
\label{fig:DainBondi}
\end{figure}

We find that our numerical results support the conjecture that on any hyperboloidal slice intersecting \scri, the inequality $\epsilon_{AB}\leq1$ holds whereby equality is obtained if and only if the data describe an (unperturbed) Kerr black hole. In figure \ref{fig:DainBondi}, we plot $(1-\epsilon_{AB})$ against the perturbation parameter $\Delta\sigma_{\cal H}$ for prescribed rotation parameters $j=0.5$ and $0.9$. Within numerical accuracy, the data confirm the statement in question, as $\epsilon_{AB}=1$ is obtained only for unperturbed data $\Delta\sigma_{\cal H}=0$\footnote{In a future work we will address the study of larger perturbations, in particular the exploration of a behavior like the one found in \cite{JarVasAns08} where $\epsilon_A$ tends back to 1 for large values of the perturbation parameter (cf.~the discussion at the end of section \ref{sec:BondiMass}). This will be of interest
in the assessment of global aspects of the rigidity conjecture.}.
  
In this manner the results provide two important tests. On the one hand, we obtain strong evidence for the validity and rigidity of the Dain-Bondi-inequality, being an {\em equality} only for the Kerr solutions. On the other hand, we have an excellent confirmation of the correctness of the numerical implementation and the results' precision which, to this extend, would be difficult to achieve by other than spectral methods. 
Note that these perturbations around the Kerr solution offer indeed a critical test in this setting, 
since any slight flaw either in the conceptual framework or in the numerical implementation could naturally spoil the non-negativity of $1-\epsilon_A$. As a particular consequence, we propose the Dain-Bondi-number as a relevant quantity in the standard picture of gravitational collapse.

\section{Discussion}
\label{sec:Discussion}
In this paper we have constructed single black hole initial data on compactified hyperboloidal slices on which the mean extrinsic curvature $K$ approaches a constant asymptotically (ACMC-condition, $K=K_0+K_4\sigma^4$). In particular, we were able to identify ACMC-slices for the Kerr solution and generated perturbed Kerr data from these. With such initial data we go a step beyond previous results, as our procedure can compute arbitrarily small deviations from equilibrium at the entire range of black hole spins. We find rapid exponential convergence of our pseudo-spectral solutions indicating analyticity of the potentials. Therefore, we believe that the stronger global CMC-condition, i.e.,~$K_4\equiv 0$, which is adopted by several authors, can be replaced in favour of the local ACMC-condition without losing the degree of smoothness. Note that the freedom provided by the ACMC-condition can be used to optimize the numerical representation. In this regard, a comparison of CMC- and ACMC-slices of the Schwarzschild metric is given in \ref{sec:ACMC_CMC_Comparison}, showing that the CMC-condition needs more coefficients for an accurate spectral approximation.

The detailed study of strong perturbations revealed a rich geometric structure regarding the appearance of different marginally trapped surfaces. Depending on the sign of the perturbation parameter $\Delta\sigma_{\cal H}$, the slices  either extend into the interior of a white hole (for sufficiently strongly negative values of $\Delta\sigma_{\cal H}$) or, if $\Delta\sigma_{\cal H}$ is sufficiently large (and positive), into the interior of a black hole, meaning that the apparent horizon is located inside the numerical grid.  A subsequent analysis, based on the evaluation of mass and angular moments of the apparent horizon, confirmed that the initial data do indeed describe a {\em perturbed} rotating Kerr black hole, whose physical properties are explored by means of the Bondi Mass $M_{\rm B}$. 

Another study was concerned with the validation of some aspects in the standard picture of gravitational collapse. In particular, we verified that our initial data satisfy the inequality\footnote{The quantities $J$ and ${\cal A}_{\cal H}$ are angular momentum and apparent horizon area respectively.}
\[{\cal A}_{{\cal H}}\le 8\pi (M_{\rm B}^2 + \sqrt{M_{\rm B}^4 - J^2})\qquad
	\mbox{(implying}\quad {\cal A}_{{\cal H}}\le 16\pi M_{\rm B}^2),\]
which gives rise to the introduction of the {\em Dain-Bondi number}
\[
\epsilon_{AB}=\frac{{\cal A}_{{\cal H}}}{8\pi (M_{\rm B}^2 + \sqrt{M_{\rm B}^4 - J^2})}
\]
as an interesting physical quantity. The results support the conjecture that the equality $\epsilon_{AB}=1$ holds true only for the (unperturbed) Kerr solutions.

The numerical solutions calculated in this work are meant to serve as initial data for a time-evolution code that solves the dynamical Einstein equations. For the development of such a code, we plan to treat the hyperboloidal evolution problem of the Einstein equations in terms of a conformally rescaled metric. In this approach we intend to utilize the {\em wave gauge} so that the Einstein equations become a system of wave equations (see e.g.~the concept suggested in \cite{Zenginoglu:2008pw}). By taking as a starting point initial data corresponding to slightly perturbed Kerr black holes, we can explore the new code in the weakly non-linear regime where the equations should be similar to corresponding linearized ones. Numerical methods to realize this project are under construction.

\section*{Acknowledgements}
We are deeply indebted to J.~L.~Jaramillo for his strong support and numerous useful hints. Also, it is a pleasure to thank L.~Buchman and A.~Zenginoğlu for many valuable discussions and N. Johnson-McDaniel for carefully reading the manuscript. This work was supported by the DFG-grant SFB/Transregio 7 ``Gravitational Wave Astronomy''.

\appendix
\section{The coordinate transformation leading to ACMC-slices}\label{sec:Appendix_A}
In section \ref{sec:ACMC} we described the coordinate transformation (\ref{eq:CoordTrafo_r}, \ref{eq:CoordTrafo_V}) which leads, for the Kerr solution, from Kerr coordinates to compactified hyperboloidal slices on which the ACMC-condition (\ref{eq:ACMC_Kerr}) is valid. The equations (\ref{eq:CoordTrafo_r}, \ref{eq:CoordTrafo_V}) contain an auxiliary function $A=A(\sigma,\mu)$ which was determined through an appropriate diagonal Pad\'e expansion. Here we provide the explicit form of this function:
\beq
A(\sigma,\mu)=\frac{\sigma Z}{N} \nonumber
\eeq
with
\begin{eqnarray*}
Z&=&24c_1\left(c_1 c_3- c_2^2\right)-\left(a_4 c_1^2+24 c_2^3-48 c_1 c_2 c_3\right) \sigma \\
N&=&24\left(c_1 c_3 - c_2^2\right)+(24 c_2 c_3-a_4 c_1) \sigma +\left(a_4 c_2-24 c_3^2\right) \sigma ^2.
\end{eqnarray*}
The coefficients $c_i=c_i(\mu\,; j)$ are given by:
\begin{eqnarray*}
c_1&=&a_1-\frac{j^2 \mu ^2}{16}\\
c_2&=&\frac{1}{8} \left(j^2-4\right)\\
c_3&=&-\frac{1}{6} \left(8+12 a_1+6 a_1^2-j^2\right)+\frac{1}{8} (1+a_1) j^2 \mu ^2-\frac{j^4 \mu ^4}{256}.
\end{eqnarray*}
To fix the free parameters $a_1$ and $a_4$ as functions of $j$, we analyse the decay of the Chebyshev coefficients $c^{(\tilde\alpha)}_k$ of the conformal lapse $\tilde{\alpha}$ (see section \ref{sec:ACMC} and footnote \ref{fn:Cheb_alpha} therein). We fit their logarithm linearly over the index, i.e. \[\text{Fit}\left[\log(M|c^{(\tilde\alpha)}_k|)\right]=f_1 + f_2\cdot k.\]
and perform a numerical minimization of $f_2$ with respect to $a_1$ and $a_4$ using a simplex algorithm. In this way we obtain the parameters corresponding to the most rapid decay. We observe that the results obtained in this manner are well represented by a simple quadratic fit,
\begin{align*}
a_1&=\frac{113}{100} + \frac{9}{500} |j| + \frac{3}{40} j^2 \\
a_4&=4!\left(\frac{161}{10} + \frac{3}{4} |j| - \frac{5}{2} j^2\right).
\end{align*}
Note the occurrence of terms containing $|j|$ which is a result of the specific quadratic fit we have chosen. In a more fundamental treatment one would try to avoid such terms, as the Kerr solution should be described by metric coefficients that are smooth with respect to $j$. However, for the purpose of this paper it is convenient to utilize the above choice,  as it realizes with little effort ACMC-slices possessing smooth functions $\tilde\alpha$ with rapidly decaying Chebyshev coefficients.

\section{Comparison of ACMC- with CMC-slices}
\label{sec:ACMC_CMC_Comparison}

In this section we perform a numerical comparison of our ACMC-slices with CMC-slices in the case of the spherically symmetric Schwarzschild black hole. The comparison consists again of analyzing  the decay of the Chebyshev coefficients $c^{(\tilde\alpha)}_k$ of the conformal lapse $\tilde{\alpha}$ in the two situations.

\begin{figure}[h!]
	\centering
	\includegraphics[clip]{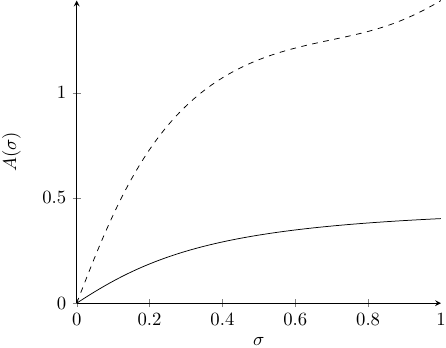}
	\includegraphics[clip]{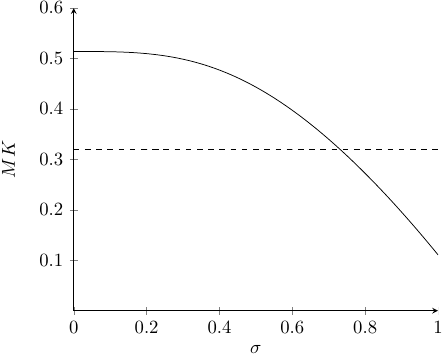}
	\caption{The quantities $A=A(\sigma)$ and $MK=MK(\sigma)$ for CMC-slices (dashed) and ACMC-slices (solid) case are plotted in the Schwarzschild case $j=0$. The slices possess asymptotically mean curvature values at \scri of $K_0\approx 0.514 M^{-1}$ for the ACMC-slicing and $K_0\approx 0.32 M^{-1}$ for the CMC-slicing. The additional parameter $C$ in the family of CMC-slices has been set to  $CM^{-2}\approx 2.88$.}
	\label{fig:Compare_CMC_ACMC_A_K}
\end{figure}

For the Schwarzschild metric we have $j=0$, and the corresponding auxiliary function $A$, appearing in the coordinate transformation leading to ACMC-slices, becomes independent of $\mu$,
\begin{eqnarray*}
A(\sigma)_{\text{\tiny{ACMC}}}=\frac{3(4478917900 \sigma^2 +1950390283\sigma)}{5178027300+14182095000\sigma+ 28592255449 \sigma^2}.
\end{eqnarray*}
A plot of this function is provided in figure \ref{fig:Compare_CMC_ACMC_A_K}. Also, the corresponding mean curvature $K_{\text{\tiny{ACMC}}}$ is displayed therein. Note that the asymptotic value of the mean curvature is [see equations (\ref{eq:a1_j}) and (\ref{eq:aCMC_K_Taylor})]
\[K_{0,\text{\tiny{ACMC}}} =\frac{3}{4M\sqrt{1+a_1}}=\frac{5}{2M}\sqrt{\frac{3}{71}}\approx 0.514 M^{-1}.\]

\begin{figure}[t!]
\centering
\includegraphics[trim=60 590 300 55,clip]{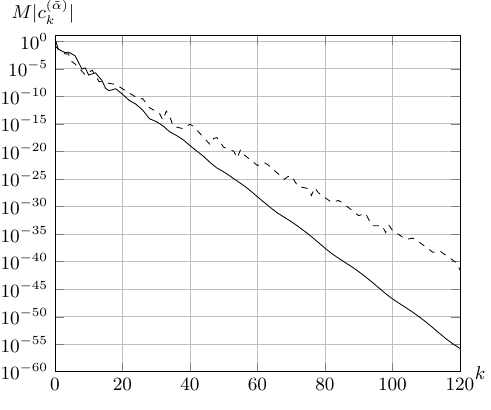}
		\caption{Chebyshev coefficients of the conformal lapse $\tilde{\alpha}$ for the Schwarzschild metric in CMC-coordinates (dashed) and ACMC-coordinates (solid). For the ACMC-slice, the order of machine precision, that is $10^{-16}$, is reached at $k=32$, while for the CMC-slice the corresponding coefficient is of the order $10^{-12}$.}
	\label{fig:Compare_CMC_ACMC}
\end{figure}

With the conformal factor $\Omega=\sigma/4M$, the associated conformal lapse turns out to be
\begin{eqnarray*}
\tilde{\alpha}_{\text{\tiny{ACMC}}}=\frac{2}{M\sqrt{(\sigma -1) \sigma ^2 A'(\sigma )^2+\left(1-2 \sigma ^2\right)A'(\sigma )+\sigma +1}}.
\end{eqnarray*}
In figure \ref{fig:Compare_CMC_ACMC}, the decay of the coefficients $M\, c^{(\tilde\alpha_{\text{\tiny{ACMC}}})}_k$ is displayed.

We now turn our attention to CMC-slices. In the Schwarzschild case, the line element (\ref{eqn:KerrCoordinates}) reduces to the Eddington-Finkelstein form, describing ingoing null rays. As described in \cite{brill:2789,Malec:2003,Malec:2009}, the coordinate transformation between the coordinates $(V,r,\vartheta,\varphi)$ and coordinates $(T,r,\vartheta,\varphi)$, in which the slices $T={\rm constant}$ satisfy the CMC-condition, 
is given by
\[T = V - \int\frac{r d r}{r-2M} - \int\frac{r a dr}{(r-2M)f},\]
where
\beq
f=\sqrt{1-\frac{2M}{r} +a^2},\qquad a=\frac{Kr}{3}-\frac{C}{r^2}
\eeq
and $K$ is the constant mean curvature. The parameter $C$ represents an additional one-parameter degree of freedom in the choice of spherically symmetric CMC-slices in the Schwarzschild metric.
For slices extending up to \scri we have 
\[C>\frac{8}{3}KM^3,\]
hence $a|_{r=2M}=-f|_{r=2M}<0$. From the relation between $T$ and $V$, we can read off the transformation leading from Eddington-Finkelstein-slices to compactified CMC-slices. Putting $T=4M\tau$, we obtain
\begin{eqnarray*}
r &=& \frac{2M}{\sigma}\\
V &=& 4 M \left(\tau + \frac{1}{\sigma} - \log(\sigma) + A_{\text{\tiny{CMC}}}(\sigma)\right)
\end{eqnarray*}
with the function $A_{\text{\tiny{CMC}}}$ given by:
\[A_{\text{\tiny{CMC}}}=\int\frac{f+a}{2\sigma^2(\sigma-1)f}d\sigma+\log(\sigma)-\frac{1}{\sigma}\]
Note that despite the apparently singular form, $A_{\text{\tiny{CMC}}}$ is {\em analytic} on the entire interval $\sigma\in[0,1]$.

Using again the conformal factor $\Omega=\sigma/4M$, the associated conformal lapse reads
\beq
\tilde{\alpha}_{\text{\tiny{CMC}}}=\frac{1}{48M}\sqrt{144(1-\sigma)\sigma^2+(8KM-3CM^{-2}\sigma^3)^2}. \nonumber
\eeq

Now, utilizing a method similar to the one described in section \ref{sec:Appendix_A}, we find that strong decay rates of the corresponding coefficients $c^{(\tilde\alpha_{\text{\tiny{CMC}}})}_k$ are realised if we put the free parameters as
\[K\approx 0.33 M^{-1},\qquad C\approx 2.88M^2.\]

In figure \ref{fig:Compare_CMC_ACMC_A_K}, the quantities $\{A_{\text{\tiny{ACMC}}}, A_{\text{\tiny{CMC}}}\}$ as well as the corresponding mean curvatures $\{K_{\text{\tiny{ACMC}}}, K_{\text{\tiny{CMC}}}\}$, obtained for these settings, are plotted. Figure \ref{fig:Compare_CMC_ACMC} shows the convergence of the Chebyshev coefficients $\{M\, c^{(\tilde\alpha_{\text{\tiny{ACMC}}})}_k,M\,c^{(\tilde\alpha_{\text{\tiny{CMC}}})}_k\}$. While the coefficients $|M\,c^{(\tilde\alpha_{\text{\tiny{ACMC}}})}_k|$ for the ACMC-slice reach $10^{-16}$, i.e.~the order of machine precision ($\sim$ 16 digits), at $k=32$, the coefficient $|M\, c^{(\tilde\alpha_{\text{\tiny{CMC}}})}_{32}|$ for the CMC-slice is only about $10^{-12}$, i.e.~larger by 4 orders of magnitude. We thus conclude that the replacement of the global CMC-condition in favour of the local ACMC-condition provides advantages in a numerical treatment.

\section*{Bibliography}
\bibliography{bibitems}
\end{document}